\documentclass[prx,twocolumn,eqsecnum,amsmath,amssymb,superscriptaddress,floatfix,english]{revtex4-1}
\usepackage{epsfig, graphicx,graphics,amsmath,amssymb,float}
\usepackage[T1]{fontenc}
\usepackage[latin9]{inputenc}
\usepackage{amsmath}
\usepackage{amssymb}

\usepackage{amscd}
\usepackage{bm}
\usepackage{psfrag}
\usepackage{bbm} 
\usepackage{babel}
\usepackage{wasysym }
\usepackage{mathrsfs}
\usepackage{color}

\definecolor{green}{RGB}{22,140,22}
\definecolor{blue}{RGB}{22,22,144}
\definecolor{reddish}{RGB}{140,22,22}

\newcommand{\be}{\begin{equation}}
\newcommand{\ee}{\end{equation}}
\newcommand{\bea}{\begin{eqnarray}}
\newcommand{\eea}{\end{eqnarray}}

\begin{document}
\title{Phonon-induced magnetoresistivity of Weyl semimetal nanowires}

\author{Alessandro De Martino}
\affiliation{Department of Mathematics, City, University of London, EC1V 0HB London, UK}

\author{Kathrin Dorn}
\affiliation{Institut f\"ur Theoretische Physik, Heinrich-Heine-Universit\"at, D-40225  D\"usseldorf, Germany}

\author{Francesco Buccheri}
\affiliation{Institut f\"ur Theoretische Physik, Heinrich-Heine-Universit\"at, D-40225  D\"usseldorf, Germany}

\author{Reinhold Egger}
\affiliation{Institut f\"ur Theoretische Physik, Heinrich-Heine-Universit\"at, D-40225  D\"usseldorf, Germany}

\begin{abstract}
We study longitudinal magnetotransport in disorder-free cylindrical Weyl semimetal nanowires.  
Our theory includes a magnetic flux  $\Phi$ piercing the nanowire and captures the finite curvature 
of the Fermi arc in the surface Brillouin zone through a boundary angle $\alpha$.  Electron backscattering 
by acoustic phonons via the deformation potential causes a finite resistivity which we evaluate by means of 
the semiclassical Boltzmann approach.   We find that low-energy transport is dominated by surface states, 
where transport observables are highly sensitive to the angle $\alpha$ and to Aharonov-Bohm phases due 
to $\Phi$. A generic subband dispersion relation allows for either one or two pairs of Fermi points. 
In the latter case, intra-node backscattering is possible and implies a parametrically larger resistivity 
than for a single Fermi point pair. As a consequence, large and abrupt resistivity changes take place across 
the transition points separating parameter regions with a different number of Fermi point pairs in a given subband.
\end{abstract}

\date{\today}

\maketitle

\section{Introduction}\label{sec1}

Weyl semimetal (WSM) materials represent one of the most intensely studied topics 
in current condensed matter physics; for recent reviews, 
see Refs.~\cite{Armitage2017,Hasan2017,Burkov2018,Lv2021,Ong2021,Zhang2021a}. 
WSM materials have pairs of Weyl nodes in the Brillouin zone which act as sources of Berry curvature, with
topological Fermi arc surface states connecting the surface projections of different Weyl nodes.  
Experimental evidence for Fermi arcs has already been accumulated for several WSM materials 
by means of surface probe techniques \cite{Hasan2017,Lv2021}, and experimental studies of 
other interesting phenomena such as the chiral anomaly \cite{Ong2021} or nonlocal Weyl orbits \cite{Zhang2021a} 
are well advanced.
Nonetheless, a satisfactory understanding of the transport properties of WSM materials is often difficult to reach
due to the intricate interplay between topological protection and backscattering mechanisms.  
In addition, it is important to include electromagnetic fields and finite size effects in specific device setups.
To give just one example, while measurements of the magnetoresistivity could in principle reveal the 
chiral anomaly \cite{Spivak2016}, the precise relation between transport observations and the chiral 
anomaly remains under intense debate \cite{Ong2021}.

In this paper, we present a theory of magnetotransport in disorder-free WSM nanowires, 
taking into account electron backscattering by acoustic phonons.  Since this device geometry 
is experimentally realizable and at the same time analytically tractable, the interplay between topological Fermi arcs, 
backscattering effects, electromagnetic fields, and finite-size effects can here be analyzed in a comprehensive manner.
The band structure and the noninteracting transport properties of clean WSM nanowires have been studied 
in Refs.~\cite{Baireuther2016,Baireuther2017,Igarashi2017,Erementchouk2018,Kaladzhyan2019,Sukhachov2020}.  
In particular, for cylindrical wires, the authors of Ref.~\cite{Kaladzhyan2019} have shown  
that the contribution of Fermi arcs to the conductance often outweighs the effect of bulk states. 
This conclusion also applies for large values of the nanowire radius, see  
Refs.~\cite{Gorbar2016,Breitkreiz2019} for related studies.  
One of the goals of this work is to quantify phonon-induced backscattering effects 
on the magnetoresistivity of WSM nanowires, in particular in parameter regions where transport
is dominated by surface states.

The importance of phonons in WSMs has been established by recent experiments  
\cite{Nguyen2020,Zhang2020,Hein2020,Osterhoudt2021}.  Phonon effects can be identified, for instance, 
through the characteristic temperature dependence of phonon-induced contributions to transport observables.
Theoretical studies of electron-phonon coupling effects have so far mainly focused on 
optical phonons and/or  phenomena unrelated to transport, see, e.g., Ref.~\cite{Rinkel2019}.
Phonon-induced backscattering effects on transport in WSMs have been studied for 
the slab geometry \cite{Resta2018} but (to the best of our knowledge)
not for nanowires.  We note that the
phonon-induced resistivity of conventional one-dimensional (1D) quantum wires 
with  parabolic (or linear) dispersion was studied by many authors
 \cite{Voit1987,Bockelmann1990,Shik1993,Mickevicius1993,Gurevich1995,Gurevich1995b,Seelig2005,Yurkevich2013}. 
 However, the dispersion relations of 1D subbands in WSM nanowires turn out to be more complex. 
For instance, a given 1D subband may allow for more than one pair of Fermi momenta.  
In such cases, new scattering processes appear which in turn directly affect the dependence of the 
resistivity on key parameters such as temperature, Fermi energy, and magnetic field.   

The consequences of this enriched complexity will here be studied for cylindrical WSM nanowires.
We employ a  two-band model describing WSMs with broken time reversal symmetry and just two bulk 
Weyl nodes \cite{Vazifeh2013,Okugawa2014,Gorbar2016,Bovenzi2018,Burrello2019}, where a boundary 
condition ensures that the current density perpendicular to the cylinder surface vanishes.
This boundary condition is parametrized by a boundary angle $\alpha$ \cite{Witten2016,Erementchouk2018}, 
where the commonly used infinite mass boundary conditions are recovered for $\alpha=0$. 
For a planar surface with $\alpha=0$, the Fermi arc curves in the surface Brillouin zone are straight lines.  
For $\alpha\ne 0$, however, one finds that Fermi arcs acquire curvature.  
By including the phenomenological parameter $\alpha$, we therefore can also address 
the case of WSM materials with curved Fermi arcs.  

We use the well-known phonon modes predicted by isotropic elastic continuum theory 
with stress-free boundary conditions in the wire geometry \cite{Landau7}, and
we assume that the deformation potential provides the dominant electron-phonon coupling.  
Including a constant magnetic field along the wire axis, we then compute the resistivity
from Boltzmann theory \cite{Landau10,Levchenko2020}.  For a complementary 
study in the context of topological insulator nanowires, see Ref.~\cite{Dorn2020}. 
In addition, we will discuss the two-terminal conductance of clean WSM nanowires 
in the zero-temperature limit, where phonon effects are frozen out.
It is interesting to compare WSM nanowires and topological insulator 
nanowires \cite{Bardarson2013,Jauregui2016}. Even though only the latter have gapped 
bulk states, we show below that surface states in both types of nanowires show a 
similar response to a magnetic flux threading the wire.
With some modifications along the lines of Ref.~\cite{Sukhachov2020}, our theory can also 
be adapted to Dirac semimetal nanowires.  Nanowires made of the Dirac semimetal material Cd$_3$As$_2$ 
have recently been synthesized; for transport experiments, see Refs.~\cite{Wang2016,Lin2017,Lin2020,Bagoyan2020,Li2021}.  
We note that first transport experiments have recently been reported for WSM nanowires as well~\cite{Nair2020,Cohn2020}.  

The paper is structured as follows. In Sec.~\ref{sec2}, we derive and discuss 
the  electronic band structure.  Assuming that the deformation potential produces 
the dominant electron-phonon coupling,  the phonon-induced resistivity has been
computed within the semiclassical Boltzmann approach as explained in Sec.~\ref{sec3}.  
Our results for transport observables are then discussed in Sec.~\ref{sec4}.
The paper concludes with a brief summary and an outlook in Sec.~\ref{sec5}.  
Details about our calculations can be found in several Appendices, and we often put $\hbar=e=c=k_{B}=1$.

\section{Electronic band structure}\label{sec2}

In this section, we address the band structure of WSM nanowires.  
In Sec.~\ref{sec2a}, we describe a two-band model for magnetic WSMs and derive the spectral 
equation for cylindrical wires. We then discuss the band structure in Sec.~\ref{sec2b}, 
in particular its dependence on magnetic flux and on the boundary angle $\alpha$.

\subsection{Model} \label{sec2a}

We start from a well-known inversion-symmetric two-band model for the single-particle 
electron states of a magnetic WSM \cite{Vazifeh2013,Okugawa2014,Gorbar2016,Bovenzi2018,Burrello2019}. 
This model describes the simplest case with just two Weyl points located at momenta 
${\bf  k}=\pm b \hat e_z$ in the Brillouin zone, where the unit vector $\hat e_z$ 
is along the $z$-direction. 
We will study a cylindrical nanowire geometry with radius $R$ and wire axis $\hat e_z$ by imposing 
a boundary condition at the cylinder surface. 
In addition, we include the effects of a constant magnetic field ${\bf B}=B\hat e_z$ 
along the wire axis, with $B>0$. We note that for a magnetic field perpendicular to the wire axis, 
 transport observables are strongly suppressed, see Ref.~\cite{Igarashi2017} for a detailed study.
 
Electronic states are then described by the low-energy model 
\cite{Vazifeh2013,Okugawa2014,Gorbar2016,Bovenzi2018,Burrello2019}
 \begin{equation}
 H_0 =v\left[ \sigma_x(-i\partial_x+A_x) + \sigma_y(-i \partial_y+A_y) \right]
 + m_k \sigma_z ,
 \label{bulkmodel}
 \end{equation}
 with the bulk Fermi velocity $v$ and Pauli matrices $\sigma_{x,y,z}$ acting in 
 a combined spin-orbital space. Clearly, the momentum $k$ along $\hat e_z$ is a good quantum
 number, and the effective mass function is given by
 \begin{equation}\label{massterm}
 m_k = \frac{v}{2b} ( k^2- b^2)  .
 \end{equation}
Throughout we focus on energies $|E|\alt vb/2$ such that the two Weyl nodes at $k=\pm b$ can be clearly
distinguished. The magnetic field is given by $B=\partial_xA_y-\partial_yA_x$, where we use the symmetric gauge, 
${\bf A}=\frac12 B (-y,x,0)$.  In units of the flux quantum $\Phi_0=hc/e$, the magnetic flux through the 
cross-section of the nanowire is encoded by the dimensionless flux parameter
\begin{equation}\label{flux}
    \Phi = \frac{\pi R^2 B}{\Phi_0}=  \frac{R^2}{2l_B^2},
\end{equation}
with the magnetic length $l_B=\sqrt{\hbar c/eB}$.
For a nanowire of radius $R=25$~nm, one finds $\Phi\approx 1$ for a magnetic field $B\approx 2$~T. 
We note that the magnetic Zeeman term has been neglected in Eq.~\eqref{bulkmodel}.  As shown in Ref.~\cite{Ramshaw2018}, even though the $g$ factor can be large in typical WSM materials, the 
Zeeman coupling is expected to cause only small quantitative changes in the band structure.  
The orbital effects of the magnetic field, on the other hand, cause qualitative differences.

Before turning to the derivation of the spectrum, let us summarize the relevant energy scales. 
First, the scale $vb/2$ corresponds to the mass gap at $k=0$, see Eq.~\eqref{massterm}.  Second,  
transverse quantization introduces the finite-size scale $v/R$.  Third, the magnetic energy scale is $v/l_B$.  
We are interested in relatively thin wires and consider low energies,
 $|E|\alt vb/2$. The number of bands in this energy range can be roughly estimated by $\sim vb/(v/R)=bR$.
Throughout this paper, we consider the case $bR\gg 1$; in concrete examples, we set $bR=10$.
Taking a typical value $b\sim 0.5~$nm$^{-1}$ in WSM materials \cite{Armitage2017,Hasan2017}, 
this choice corresponds to a nanowire radius $R\sim 20$~nm.  The ratio between the magnetic scale $v/l_B$ 
and the finite-size scale $v/R$ remains as free parameter determined by $\Phi$.

\begin{figure*}
\includegraphics[width=5.9cm]{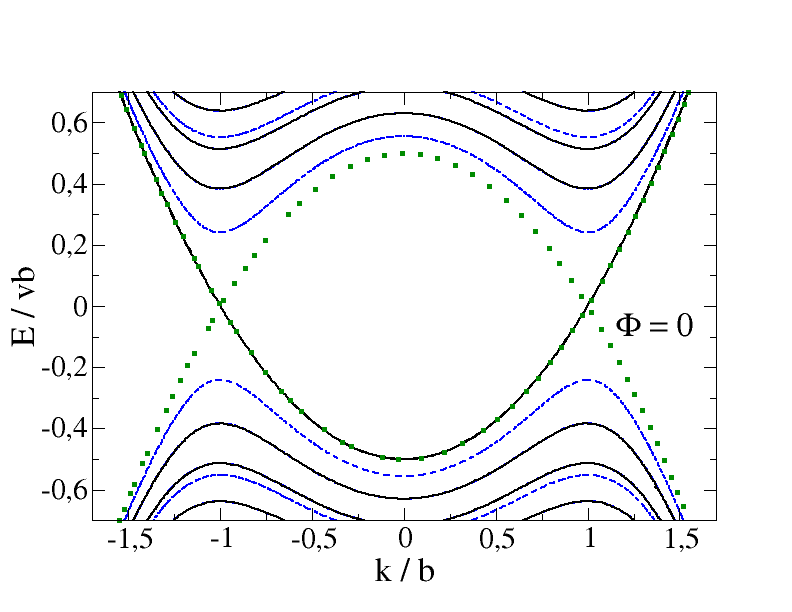}
\includegraphics[width=5.9cm]{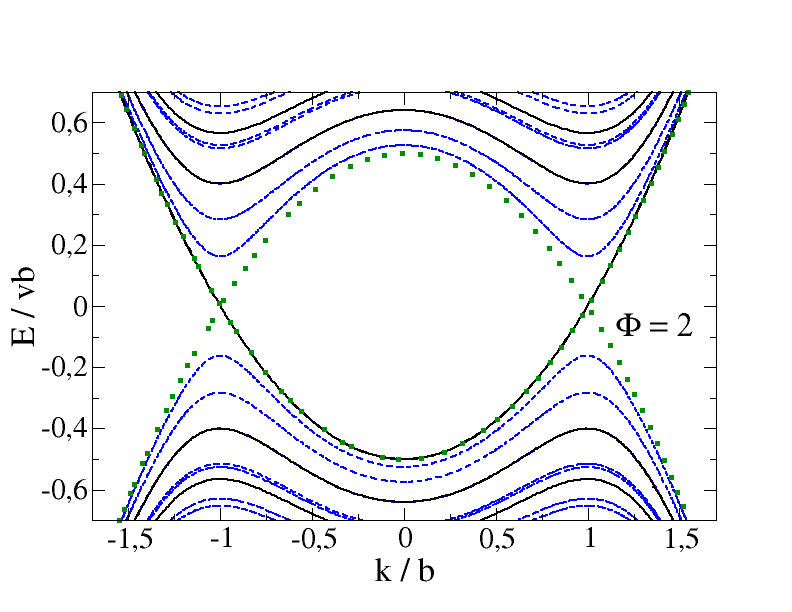}
\includegraphics[width=5.9cm]{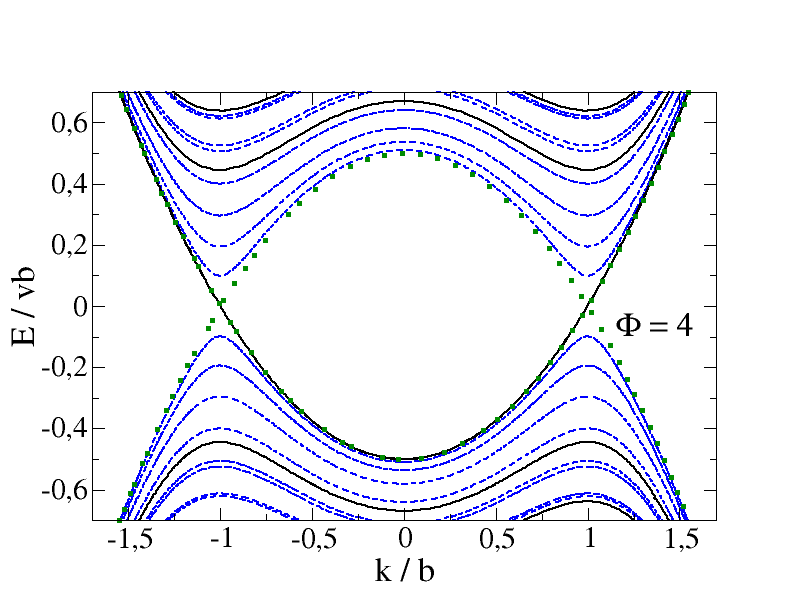}
\caption{Energy bands $E_{k,j,p}$ vs momentum $k$ for a WSM nanowire with $\alpha = \pi/2$, $bR = 10$, and 
magnetic flux parameter $\Phi=0$ (left), $\Phi=2$ (center), and $\Phi=4$ (right panel), see Eq.~\eqref{bands}.
Dashed blue (solid black) curves represent $j < 0$ ($j>0$) states. 
For this value of $\alpha$, the Fermi arc surface states with $E_{k,j>0}=m_k$ are degenerate.
For all other bands, we find states with $-21/2\le j\le 27/2$ in the shown energy range.
Green dotted curves show $E=\pm m_k$. }
\label{fig1}
\end{figure*}

We proceed by employing polar coordinates, $(x,y)=r(\cos\phi,\sin\phi)$, 
with unit vectors $\hat e_r$ and $\hat e_\phi$.
Below we will also use the dimensionless radial variable
\begin{equation}\label{xidef}
\xi=\frac{r^2}{2l_B^2}, \quad {\rm i.e.,} \ \ \xi/\Phi = (r/R)^2.
\end{equation}
From Eq.~\eqref{bulkmodel} one then finds that the angular momentum operator 
$J_z=-i\partial_\phi + \frac12\sigma_z$ with half-integer eigenvalues $j$ is conserved. 
Spinor eigenfunctions are thus given by 
\begin{equation}\label{general}
    \Psi_{k,j}({\bf r}) = \frac{e^{ik z}}{\sqrt{L}} 
    \frac{e^{i j\phi}}{\sqrt{2\pi}} \left(\begin{array}{c} 
    e^{-i\phi/2}\, Y_+(\xi) \\ ie^{i\phi/2}\, Y_-(\xi) \end{array}\right),
\end{equation}
where the wire length $L$ appears for normalization.  
The real-valued radial eigenfunctions $Y_\pm(\xi)$ are combined to form radial spinors,
\begin{equation}\label{PHidef}
    Y_{k,j}(\xi) = \left( \begin{array}{c} Y_+(\xi) \\ Y_-(\xi) \end{array}\right),
    \quad   l_B^2\int_0^{\Phi} d\xi \,( Y_+^2+Y_-^2) = 1,
\end{equation} 
where the normalization condition has been adapted to the cylindrical geometry.
Using Eqs.~\eqref{general} and \eqref{PHidef}, $H_0\Psi=E\Psi$ 
reduces to the radial equation 
\begin{equation}\label{spinoransatz}
 \left( \begin{array}{cc} - {\cal E}_-  &  \sqrt{\xi} \partial_\xi + \frac{\xi+j+\frac{1}{2}}{2\sqrt{\xi}} \\
-\sqrt{\xi} \partial_\xi + \frac{\xi+j-\frac{1}{2}}{2\sqrt{\xi}}  &-{\cal E}_+\end{array}
 \right) Y_{k,j}(\xi) =0,
 \end{equation}
with the dimensionless quantities
\begin{equation}\label{epm}
    {\cal E}_\pm(k,E) =\frac{E\pm m_k}{\sqrt2 v/l_B} .
\end{equation}
We require regularity of $Y(\xi)$ at the origin $\xi=0$.  
Then the general solution of Eq.~\eqref{spinoransatz} 
can be expressed in terms of the confluent hypergeometric function $M(a,b;\xi)$ \cite{NIST}.
Using the notation
\begin{equation}\label{adef}
    a_j = \left(j+1/2\right)\Theta(j) -{\cal E}_+ {\cal E}_- ,
\end{equation}
with the Heaviside step function $\Theta$ and keeping the dependence on $k$ and $E$ implicit,
we obtain (up to normalization)
 \begin{widetext}
\begin{equation}\label{psiplus}
    Y_{k,j}(\xi) = \left\{ \begin{array}{cc}  \xi^{\frac12 (j-\frac12)} e^{-\xi/2}
    \left( \begin{array}{c} (j+\frac12) \, M(a_j,j+\frac12;\xi)  \\
    {\cal E}_-\sqrt{\xi} \, M(a_j,j+\frac{3}{2};\xi)\end{array}\right),& j>0,\\ & \\
 \xi^{-\frac12 (j+\frac12)} e^{-\xi/2}
    \left( \begin{array}{c} {\cal E}_+\sqrt{\xi} \, M(a_j+1,-j+\frac{3}{2};\xi)  \\
(j-\frac12)\, M(a_j,-j+\frac{1}{2};\xi)\end{array}\right), &j<0. \end{array}\right .
\end{equation}
 \end{widetext}

\begin{figure*}
\includegraphics[width=5.9cm]{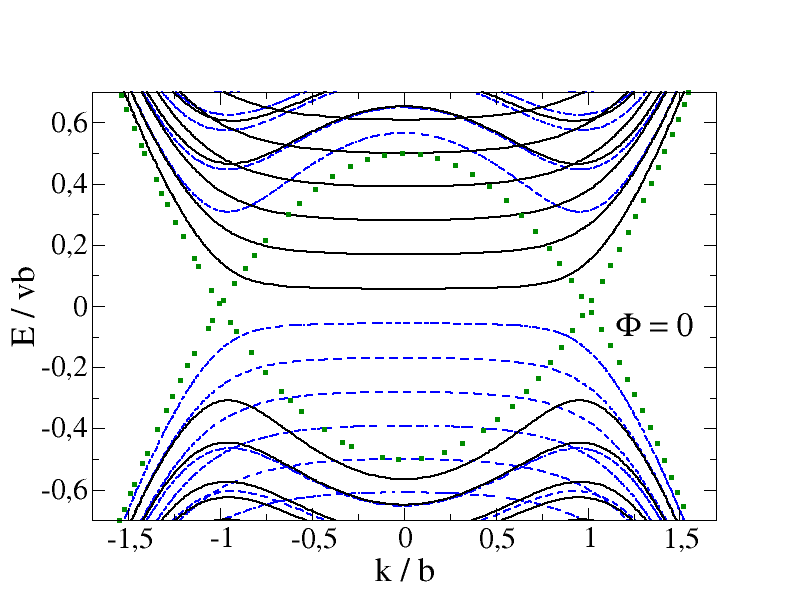}
\includegraphics[width=5.9cm]{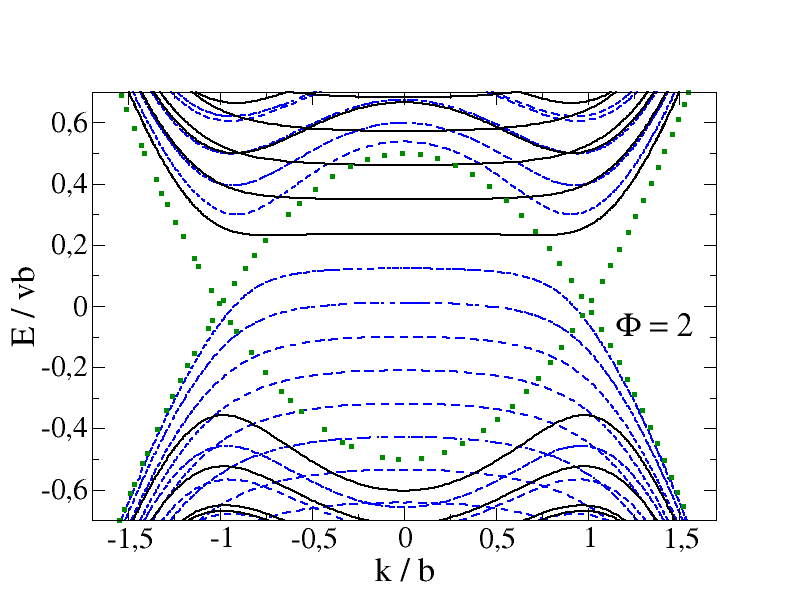}
\includegraphics[width=5.9cm]{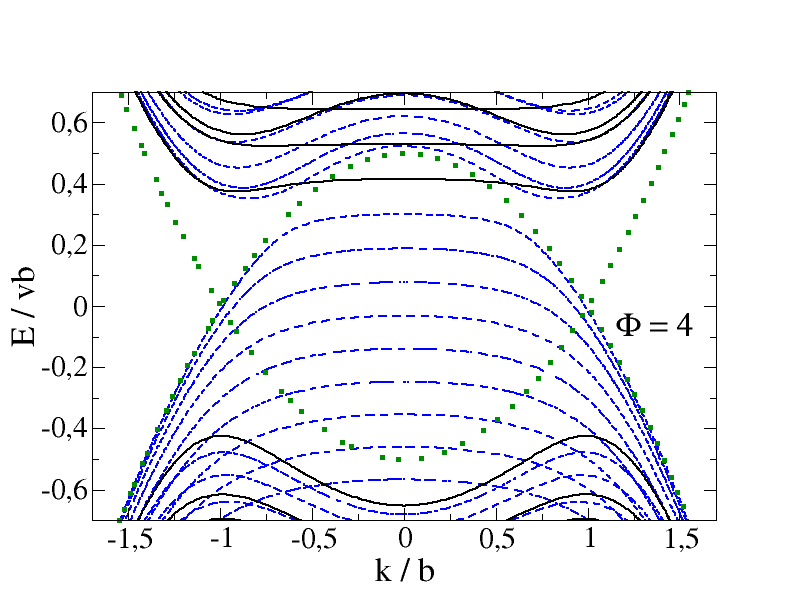}
\caption{Energy bands $E_{k,j,p}$ vs momentum $k$ for  $\alpha=0$ and several  $\Phi$. 
All other parameters and conventions are as in Fig.~\ref{fig1}.
}
\label{fig2}
\end{figure*}
\begin{figure*}
\includegraphics[width=5.9cm]{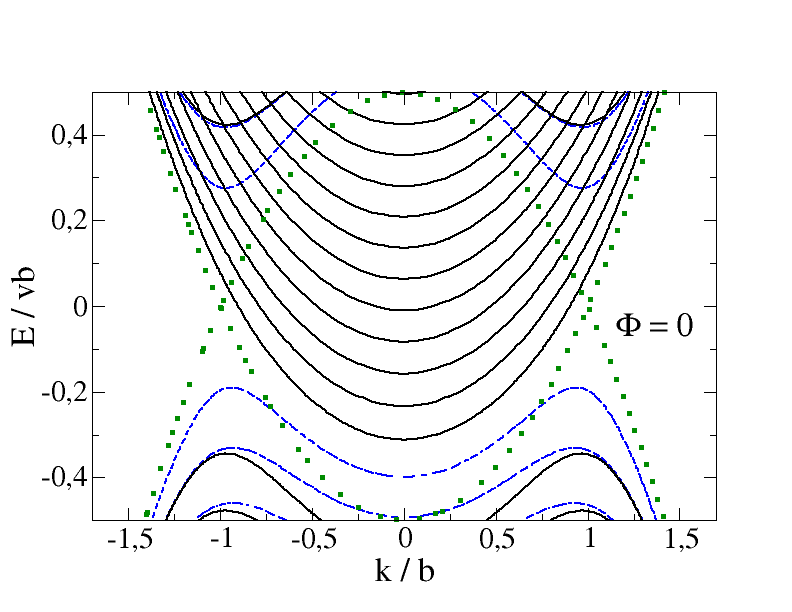}
\includegraphics[width=5.9cm]{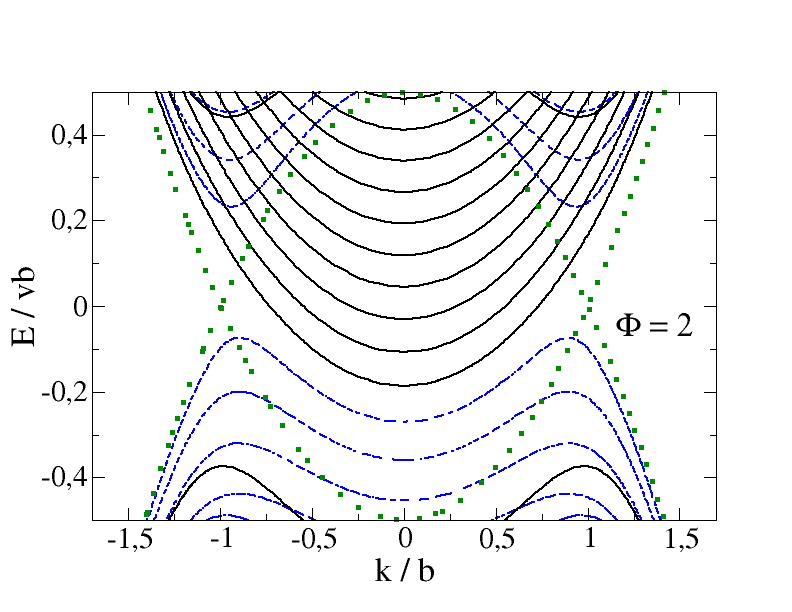}
\includegraphics[width=5.9cm]{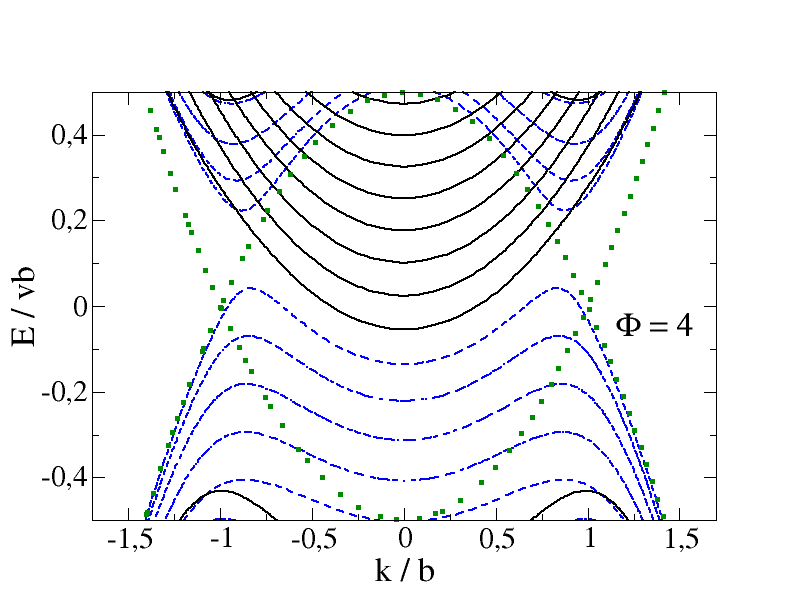}
\caption{Energy dispersion $E_{k,j,p}$ vs $k$ for $\alpha=\pi/4$ and several $\Phi$.
All other parameters and conventions are as in Fig.~\ref{fig1}.}
\label{fig3}
\end{figure*}

The finite cylinder radius $R$ now enters through a boundary condition at the surface
$r=R$, i.e., for $\xi=\Phi$.
Following Refs.~\cite{Witten2016,Erementchouk2018}, 
this boundary condition is written in the form
\begin{equation}
    \label{bc}
    M(\alpha)\Psi(R) = \pm \Psi(R),\quad M(\alpha) = \sigma_\phi \cos\alpha + \sigma_z \sin\alpha, 
\end{equation}
with $\sigma_\phi=e^{-i\frac{\phi}{2}\sigma_z} \sigma_y e^{i\frac{\phi}{2}\sigma_z}$.
We consider the $+1$ eigenvalue in Eq.~\eqref{bc} for $-\pi/2<\alpha\le \pi/2$ in what follows.
The boundary condition \eqref{bc} imposes that on the surface of the wire
the pseudospin direction lies in the tangent plane, at an angle $\alpha$ 
with respect to the circumferential direction $\hat e_\phi$. Importantly,
this condition preserves angular momentum conservation and 
ensures a vanishing local current density through the surface. 
This last condition is the same one would impose on a conventional 
semiconducting nanowire, but the form of the effective Hamiltonian in a WSM allows for one free parameter, 
the boundary angle $\alpha$. This is a non-universal parameter which 
in general will depend on both the WSM material and the precise surface structure.

Using Eq.~\eqref{general} to express $\Psi$ in terms of radial functions,
Eq.~\eqref{bc} is equivalently written as
\begin{equation}\label{bc2}
\frac{Y_+(\Phi)}{Y_-(\Phi)} =\tan\left(\frac{\alpha}{2}+\frac{\pi}{4}\right) .
\end{equation}
The choice $\alpha=0$ implements infinite mass boundary conditions
\cite{Igarashi2017,Kaladzhyan2019}, defined by a $\xi$-dependent mass given by $m_k$ in 
Eq.~\eqref{massterm} for $\xi<\Phi$ but $m_k\to \infty$ for $\xi>\Phi$. 

\subsection{Band structure}\label{sec2b}

The solutions admitted by the boundary condition~\eqref{bc2} determine the energy 
spectrum of the nanowire, which consists of 1D subbands labeled by the angular 
momentum $j$ and a radial band index $p$. 
By inversion symmetry, the respective subband dispersion $\varepsilon_k\equiv E_{k,j,p}$ is always symmetric, $\varepsilon_{-k}=
\varepsilon_k$.  The qualitative features of the spectrum depend on the interplay of the three dimensionless parameters  $bR, \Phi$,
and $\alpha$ characterizing our system.  

\begin{figure}
\includegraphics[width=0.45\textwidth]{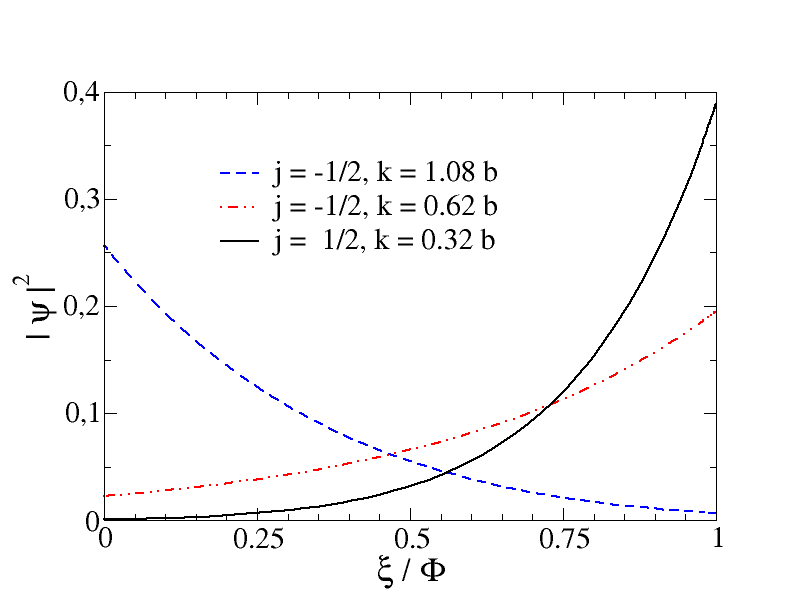}
\caption{Probability density $|\Psi_{k,j}|^2$ vs radial coordinate $\xi/\Phi=(r/R)^2$ for three 
eigenstates with energy $E=-0.15vb$, using $\alpha=\pi/4$, $bR=10$, and $\Phi=2$, 
see central panel in Fig.~\ref{fig3}. The case $k=0.32b$ and $j=1/2$ corresponds to a Fermi arc state. 
For the $j=-1/2$ subband, we find a bulk state at $k=1.08b$ but a surface state at $k=0.62b$. 
}
\label{fig4}
\end{figure}

In general, the spectral condition \eqref{bc2} has to be solved numerically, but in several limiting cases,
analytical progress is possible.  In particular,  an  approximate solution for the dispersion of Fermi arc surface states
will be given below. The full spectrum can be obtained in closed form for the boundary angle $\alpha=\pi/2$, 
see App.~\ref{appA}, and is illustrated in Fig.~\ref{fig1} for several values of the magnetic flux 
parameter $\Phi$. For all angular momenta $j>0$, we obtain degenerate Fermi arc surface states with 
the $\Phi$-independent dispersion relation $\varepsilon_k=m_k$.  
However, the point $\alpha=\pi/2$ is quite special since for $\alpha<\pi/2$, we will see below that
the Fermi arc degeneracy is lifted and the arc dispersion depends on $\Phi$. 
To illustrate the typical band structure found for $\alpha<\pi/2$,  results obtained by numerical
solution of Eq.~\eqref{bc2} are shown for $\alpha=0$ in Fig.~\ref{fig2}, and for $\alpha=\pi/4$ in Fig.~\ref{fig3}.
The radial probability density distribution is shown for selected states in Fig.~\ref{fig4}.

In order to better understand the band structure, we next discuss surface states. 
As we show in App.~\ref{appB}, the radial Dirac-Weyl equation \eqref{spinoransatz} admits solutions 
where the radial spinor wave function is localized at the surface, $Y(r)\propto e^{-\kappa(R-r)}Y(R)$. 
The inverse decay length must satisfy $\kappa R\gg 1$ to describe a proper surface state and follows as
\begin{equation}\label{kappadef}
\kappa = \sqrt{  \frac{(j+\Phi)^2}{R^2}+\frac{m_k^2-E^2}{v^2} },
\end{equation}
where the surface state dispersion is given by 
\begin{equation}\label{fermiarc}
E_{k,j}=  \frac{v (j+\Phi)} {R} \cos\alpha+m_k\sin\alpha
\end{equation}
under the condition
\begin{equation}\label{cond1}
\frac{v(j+\Phi)}{R} \sin\alpha-m_k\cos \alpha >0. 
\end{equation}
Equations \eqref{fermiarc} and \eqref{cond1} describe Fermi arc states in WSM nanowires in the presence of a 
magnetic flux threading the wire. This flux enters only through the shift $j\to j+\Phi$, just as for the 
surface states in topological insulator nanowires \cite{Bardarson2013,Dorn2020}.  
In the absence of a magnetic field and for very large $R$,  
Eq.~\eqref{fermiarc} reproduces the known Fermi arc dispersion for a planar surface~\cite{Burrello2019}. 
The approximations leading to Eqs.~\eqref{fermiarc} and \eqref{cond1}, see App.~\ref{appB}, hold under the condition 
\begin{equation}\label{cond2}
    \left|\frac{j -\Phi}{j +\Phi}\right| \ll \kappa R.
\end{equation}
We observe that for $\kappa R\gg 1$, Eq.~\eqref{cond2} is always fulfilled except for nearly half-integer  
values of $\Phi$, where the subband with the angular momentum $j$ closest to $-\Phi$ 
can violate Eq.~\eqref{cond2}.

A comparison to the numerical solution of Eq.~\eqref{bc2} shows that under the above conditions,
the dispersion of Fermi arc states in cylindrical WSM nanowires is well approximated by 
Eq.~\eqref{fermiarc}, see App.~\ref{appB}.
For $\alpha=0$, the spectrum in Fig.~\ref{fig2} exhibits a sequence of almost flat Fermi arc states for $-b< k<b$,  
with  energy spacing given by the finize-size scale $v/R$. This numerical result is
in accordance with Eq.~\eqref{fermiarc}.  For finite $\alpha$, the bands disperse.
This case is illustrated for $\alpha=\pi/4$ in Fig.~\ref{fig3}, where the Fermi arc dispersion again 
agrees with Eq.~\eqref{fermiarc}.  Apart from an increase in radial probability density as the surface is approached, see
Fig.~\ref{fig4}, surface states can therefore also be identified by a strong 
sensitivity of the dispersion to the boundary angle $\alpha$.

Next we turn to bulk states, where the probability density is large away from the surface.
For $R\to \infty$, Landau states follow by standard steps
from the expressions in Sec.~\ref{sec2a}. Using the magnetic length $l_B=\sqrt{\hbar c/eB}$ 
and the index $n=0,1,2,\ldots$, their dispersion is given by
\begin{equation}\label{bulk}
    E_{k,j,p}= \left\{ \begin{array}{cc} \pm \sqrt{\frac{ 2(n+j+\frac12) v^2}{l_B^2}+ m_k^2 },& j>0, \, p=(n,\pm),\\
    \pm \sqrt{\frac{ 2n v^2}{l_B^2}+ m_k^2 },& j<0,\, p=(n\ge 1,\pm),\\
    - m_k, & j<0,\, p=n=0.\end{array}\right.
\end{equation}
The states with $j<0$ and $n=0$ are chiral zero modes \cite{Armitage2017}.  
For a finite radius $R$, these bulk dispersions are obtained as long as $l_B\ll R$ and the 
corresponding wave functions are centered within the nanowire, far from the surface. 
For a given Landau level, upon decreasing $j$, the states have increasing weight
near the surface and eventually become chiral edge states. 
In general, surface states can thus represent Fermi arc or 
chiral edge states. By monitoring the magnetic field dependence, 
the character of a given surface state can be revealed, as
only Fermi arcs remain well-defined surface states for $B\to 0$. 

We finally note that in the finite-size geometry considered here, there is not a sharp
distinction between surface bands and bulk bands. 
The character of the states (bulk vs surface) within a given subband depends on $k$. 
This is illustrated in Fig.~\ref{fig4}, where we show the radial profile of the
probability amplitude for states with energy $E= - 0.15vb$ in bands with $j=\pm 1/2$ as an example. 
The probability density mainly accumulates near the surface for the 
state with $j=1/2$. However, for the two states in the $j=-1/2$ subband, which correspond to opposite sides of 
the extremum in the dispersion at $k\approx b$, we observe that one is a bulk state and the
other a surface state. Specifically, in Fig.~\ref{fig4}, the $j=-1/2$ state with $k=1.08b$ has a large probability density near the center of the nanowire (bulk state), while the state with $k=0.62b$ is peaked near its boundary (surface state).

\section{Phonon-induced resistivity: Boltzmann theory}\label{sec3}

In this section, we derive the phonon-induced resistivity in WSM nanowires with 
the band structure described in Sec.~\ref{sec2}.  
Our model for including electron-phonon scattering effects is summarized in Sec.~\ref{sec3a}.
We compute the longitudinal magnetoresistivity, $\rho=\rho(T,\mu,\Phi,\alpha)$,  
 in the linear response regime from semiclassical Boltzmann theory \cite{Landau10,Levchenko2020}, see 
Sec.~\ref{sec3b}.   We  separately consider the resistivity contributions from
bands with a single pair of Fermi points, see Sec.~\ref{sec3c}, and from bands with 
two pairs of Fermi points, see Sec.~\ref{sec3d}.  

\subsection{Electron-phonon coupling}\label{sec3a}

We first describe the effects of a deformation potential coupling between phonons 
and electrons at low energy scales, where we include only acoustic phonon modes 
that are able to generate such a coupling.  
Experiments on WSM nanowires are often carried out on nanowires deposited on a 
substrate (see, e.g., \cite{Wang2016,Lin2017,Lin2020}), and we here focus 
on phonon modes which remain gapless even in the presence of a substrate.
Since the flexural (bending) modes with finite angular momentum are expected to be gapped,
in what follows we only take into account the longitudinal acoustic phonon 
mode with zero angular momentum and dispersion $\omega_q=c_L |q|$, 
where the sound velocity $c_L$ is typically small against the
Fermi velocity $v$ and the phonon momentum $q$ is defined along $\hat e_z$.
Using typical parameters for $c_L$ and $v$ in the WSM material TaAs \cite{Peng2016} 
for an order of magnitude estimate, we find $c_L/v\sim 0.01$.  
The phonon momenta $q$ responsible for 
low-temperature backscattering processes then satisfy $qR\ll 1$ and correspond to 
effectively 1D phonon modes.

We assume an isotropic elastic continuum model with 
stress-free boundary conditions at the cylinder surface \cite{Landau7}.  
The resulting phonon modes are well known.  In contrast to most previous works, where phonon backscattering in 
1D wires has been examined for three-dimensional phonon modes, 
we focus on the 1D phonon mode corresponding to longitudinal acoustic phonons with zero angular momentum.
With the bosonic annihilation operators $b_q$, the bulk mass density $\rho_M$, and Poisson's ratio $\nu$ (where $0<\nu<1/2$), 
the displacement field operator is then given by \cite{Landau7,Dorn2020}
\begin{equation}\label{displa}
    {\bf u}({\bf r}) = \int \frac{dq}{2\pi} \frac{{\rm sgn}(q) e^{iqz}}{\sqrt{2\pi R^2 \rho_M  \omega_q}}  \left( \nu 
    qr \hat e_r+i\hat e_z\right) \left [ b_q^{}+b_{-q}^\dagger\right ] .
\end{equation}
Assuming that the deformation potential is the dominant coupling mechanism, 
the electron-phonon interaction reads
\begin{equation}\label{defpot}
    H_{\rm ep} = g_0 \int d{\bf r}\, \rho_e({\bf r}) \nabla\cdot {\bf u}({\bf r}) ,
\end{equation}
where the coupling constant $g_0$ has dimension of energy and $\rho_e({\bf r})$ is the 
electron density operator.  
Unfortunately, it is hard to get reliable theoretical predictions for the value of $g_0$ 
since this coupling constant is strongly affected by screening processes.  
A standard Thomas-Fermi argument predicts $g_0\propto 1/n_b(\mu)$, where $n_b(\varepsilon)$
is the bulk density of states. Since the latter vanishes for chemical potential $\mu\to 0$, 
we expect large couplings for $|\mu|\ll vb$.
Recent experimental results suggest that  the 
electron-phonon coupling is of the order of $10$ meV but varies substantially in a small energy range \cite{shojaei2021}.
In any case, the value of $g_0$ affects the phonon-induced resistivity only via the overall 
resistivity scale $\rho_0$ discussed below. 

We then express the electronic density $\rho_e({\bf r})$ in terms of the normalized radial 
eigenstates $Y_{k,j,p}(\xi)$ in Eq.~\eqref{PHidef}, 
with fermion annihilation operators $c_{k,j,p}$.  Using Eq.~\eqref{displa} and taking the 
limit $L\to \infty$, we obtain
\begin{eqnarray}\nonumber
 H_{\rm ep} &=& -(1-2\nu)g_0 \sum_{j,p,p'} \int \frac{dk}{2\pi}\frac{dk'}{2\pi} \frac{dq}{2\pi} \delta(k-k'-q) \\
 &\times& \frac{|q|R l_B^2}{\sqrt{2\pi\rho_M\omega_q}} \nonumber
 \int_0^{\Phi} d\xi \,\, Y_{k',j,p'}^\dagger(\xi) \cdot Y_{k,j,p}(\xi)  \\ \label{Hep}
 &\times& \left( b_q^{}+b^\dagger_{-q}\right)  c_{k',j,p'}^\dagger c_{k,j,p}^{}.
\end{eqnarray}
Since we include only longitudinal acoustic phonons with zero angular momentum, the
electron-phonon interaction~\eqref{Hep} only couples electronic states with the same angular momentum $j$.
In principle, scattering processes between different radial eigenmodes with the same $j$ are possible.  However,
we here focus on parameter regions where at most a single radial band for given $j$ crosses the Fermi level. 
This simplification is justified for relatively thin nanowires at low energies, $|\mu|\alt vb/2$. 
(We have explicitly verified this point by monitoring the 
band structure for all results presented in this work.)
We note that in order to describe the resistivity in the ultimate bulk  
limit $bR\to \infty$, arbitrary scattering processes involving different radial 
modes with the same $j$ become relevant. This problem is, however, beyond the scope of this work.

\subsection{Boltzmann theory}\label{sec3b}

For a translationally invariant nanowire in a weak constant electric field $E\hat e_z$, 
Ohm's law states that a steady-state charge current density $J\hat e_z$ with $J=\sigma E$ will flow.   
In the Boltzmann approach, one uses transition rates obtained from Fermi's golden rule to compute the linear conductivity $\sigma$ \cite{Landau10}; the
resistivity then follows as $\rho=1/\sigma$. On this perturbative level, electron-phonon scattering processes 
generated by $H_{\rm ep}$ always scatter an initial electronic state with angular momentum $j$ to a final
state with the same angular momentum.  Ohm's law then implies that the 
conductivity contributions $\sigma_j=1/\rho_j$ from different angular momentum  channels simply add up,  
\begin{equation}\label{rho1}
    \frac{1}{\rho} = \sum_j \frac{1}{\rho_j},
\end{equation}
and we only have to tackle the problem for fixed angular momentum $j$.   
However, in cases where processes beyond Fermi's golden rule become important, 
Eq.~\eqref{rho1} represents an approximation.  

We obtain the resistivity contribution $\rho_j$  by 
solving a linearized Boltzmann equation for the 1D subband with angular momentum $j$.   
We use the notation $\varepsilon_k=E_{k,j,p}=\varepsilon_{-k}$ and $Y_k=Y_{k,j,p}$, 
and as discussed in Sec.~\ref{sec3a}, we focus on  parameter regions with a single 
radial band for given $j$. The steady-state distribution function is then written  as 
\begin{equation}
    n_k = n_{F}(\varepsilon_k)+ \delta n_k, \quad n_{F}(\varepsilon)=\frac{1}{e^{\beta(\varepsilon-\mu)}+1},
\end{equation} 
where  $\delta n_k$ is the nonequilibrium correction to the Fermi equilibrium 
distribution and $\beta=1/T$. We follow standard practice and parametrize $\delta n_k$ 
by a function $g(\varepsilon_k)$ \cite{Landau10},
\begin{equation}
    \delta n_k= -eE \left(-\frac{\partial n_{F}(\varepsilon_k)}{\partial\varepsilon_k}\right) v_k g(\varepsilon_k), \quad v_k=\partial_k\varepsilon_k. 
\end{equation}
With $\omega_q=c_L|q|$ and following the notation of Ref.~\cite{Levchenko2020},
the linearized Boltzmann equation can be written as 
\begin{eqnarray}\nonumber
 v_k\frac{\partial n_{F}(\varepsilon_k)}{\partial\varepsilon_k} &=& 
 \frac{1}{T} \int_{-\infty}^\infty \frac{dk'}{2\pi} 
 D(k,k') \left[ v_{k'}g(\varepsilon_{k'})-v_k g(\varepsilon_k) \right] \\ &\times & \label{linbol}
 \sum_{\nu=\pm} \delta\left( \varepsilon_k-\varepsilon_{k'}-\nu  \omega_{k-k'}\right),
\end{eqnarray}
 with the symmetric kernel 
\begin{equation}
    D(k,k') = W(k,k') \frac{n_{F}(\varepsilon_k) n_{F}(\varepsilon_{k'})}{
    \left| e^{-\beta(\varepsilon_k-\mu)} -e^{-\beta(\varepsilon_{k'}-\mu)} \right|}.
\end{equation}
Here $W(k',k)$ denotes the transition probability for scattering from an initial state with an electron with momentum $k$ to
a final state with an electron with momentum $k'$ under emission of a phonon with momentum $q=k-k'$. 
Microreversibility dictates that the same probability also describes the phonon absorption process \cite{Levchenko2020,Landau10}, 
where the initial state contains an electron with momentum $k$ and a phonon with momentum $q=k'-k$, 
and the final state has an electron with momentum $k'$.  We thus have $W(k,k')=W(k',k)$.  

For the electron-phonon interaction \eqref{Hep}, Fermi's golden rule yields
\begin{equation}
    W(k,k') =  2\pi Z v^2 |k-k'|\, {\cal I}_{k,k'}  ,  \label{Wdef}
\end{equation}
with (squared) dimensionless overlap integrals,
\begin{equation}\label{overlap}
  {\cal I}_{k',k}= {\cal I}_{k,k'}= \left| l_B^2 \int_0^{\Phi} d\xi \, Y_{k'}^\dagger(\xi) \cdot Y_{k}(\xi)\right|^2 ,
\end{equation}
and the dimensionless electron-phonon coupling parameter
\begin{equation}\label{Zdef}
    Z= \frac{g_0^2 (1-2\nu)^2}{2\pi \hbar R^2 \rho_M c_L v^2 } .
\end{equation}
For an order-of-magnitude estimate, we assume $g_0(1-2\nu)\sim 1$~eV and 
consider TaAs material parameters with $\rho_M \approx 10$~g$/$cm$^3$, $c_L \approx 2000$~m$/$s, and $c_L/v\sim 0.01$.  
For a nanowire with radius $R\sim 20$~nm, Eq.~\eqref{Zdef} then gives $Z\sim 10^{-8}$. 

Once the solution to Eq.~\eqref{linbol} has been determined, the resistivity follows from 
\begin{equation}\label{conductivity}
\frac{1}{\rho_j} = e^2 \int \frac{dk}{2\pi} \, v_k^2 
     \left( -\frac{\partial n_{F}(\varepsilon_k)}{\partial\varepsilon_k}\right) g(\varepsilon_k).
\end{equation}
The linearized Boltzmann equation \eqref{linbol} can be solved by a constant function $g(\varepsilon)=g$. Following \cite{Levchenko2020}, we find 
\begin{eqnarray}\nonumber
    g &=& \frac{C}{A},\quad 
    C =  \int \frac{dk}{2\pi} v_k^2 \left( -\frac{\partial n_{F}(\varepsilon_k)}{\partial\varepsilon_k}\right),\\
    \nonumber
    A&=& \frac{1}{2T}\int \frac{dk}{2\pi} \frac{dk'}{2\pi} D(k',k) \left(v_{k'}-v_{k}\right)^2\times \\
&\times& \label{CB}
    \sum_{\nu=\pm} \delta\left (\varepsilon_k-\varepsilon_{k'}-\nu \omega_{k-k'}\right).
    \end{eqnarray}
Since the linearized Boltzmann equation is a non-singular linear integral equation, it has a unique solution.  
Within the validity regime of the approximations made above, Eq.~\eqref{CB} therefore describes the only solution.    
    
Below we separately consider subbands with one or two local extrema (dubbed ``valleys'' or ``nodes'').  Both single-valley and two-valley subbands appear in the spectrum of WSM nanowires, see Sec.~\ref{sec2b}.
Single-valley subbands have a local extremum at $k=0$ and closely resemble the dispersion encountered in 
conventional 1D quantum wires with a single pair of Fermi points, $k=\pm k_F$.
Two-valley subbands instead have local extrema near $k\approx\pm b$, giving rise to
 a regular or inverted mexican hat shape of the dispersion.  In that case, the number of Fermi point pairs
 (one or two) depends on the chemical potential.

\subsection{One pair of Fermi points} \label{sec3c}

We first consider the case characterized by a single pair of Fermi points 
at $k=\pm k_F$ (with $k_F>0$), where  
the Fermi velocity is given by $v_F = |\partial_k \varepsilon_{k=k_F}|$.  
We consider low temperatures and assume that typical phonon energies are much smaller
than the relevant electron energies $\varepsilon_{k}$ and $\varepsilon_{k'}$ in Eq.~\eqref{CB},
i.e., the latter energies are very close to the Fermi energy $\mu=\varepsilon_{\pm k_F}$.
The integration over momenta in Eq.~\eqref{CB} is then limited to a small region around the Fermi momenta, and 
we can linearize the dispersion for $k\approx \pm k_F$. The linearization 
breaks down near the band bottom (or when approaching the transition to a regime with 
two pairs of Fermi points in a two-valley subband), where the respective resistivity contribution 
may formally diverge. However, as long as other bands with finite resistivity remain present, no contribution to the total 
resistivity~\eqref{rho1} arises from such a divergence.

As  detailed in App.~\ref{appC}, from Eq.~\eqref{CB}  we then find $C\simeq v_F/\pi$ and  
\begin{equation}\label{afinal}
    A \simeq \frac{4k_F}{\pi} Z v^2 \, {\cal F} \left( T_{\rm BG}/T \right),
\end{equation}
where we use the function 
\begin{equation}\label{auxfunc}
    {\cal F}(X) = \frac{X/2}{\sinh^2(X/2)}.
\end{equation}
The Bloch-Gr\"uneisen temperature is defined by
\begin{equation}\label{TBG}
    T_{\rm BG} = \omega_{2k_F} =  2 c_L k_F.
\end{equation}
To give a typical order of magnitude, for $k_F\sim b$ and TaAs parameters, we find $T_{\rm BG}\sim 10$~K.
Since only phonons with momentum $q\sim 2k_F$ can efficiently backscatter electrons, 
phonons with energy $\sim T_{\rm BG}$ are required in such $2k_F$ processes. 
From Eq.~\eqref{conductivity}, we then find 
\begin{equation}
    \rho_j \simeq  \frac{\pi \hbar}{e^2 v_F} \frac{A}{C}.
\end{equation}
With the overall resistivity scale 
\begin{equation}\label{rho0}
\rho_0= \frac{h}{e^2} Zb ,
\end{equation}  
we thus arrive at 
\begin{equation}\label{single}
    \frac{\rho_j}{\rho_0}  = \frac{2k_F}{b} \frac{v^2}{v_F^2} \, {\cal F}\left(T_{\rm BG}/T\right).
\end{equation}
We emphasize that both $v_F$ and $k_F$, and therefore also $T_{\rm BG}$, 
depend on the angular momentum $j$. These quantities can be obtained numerically 
from the band structure discussed in Sec.~\ref{sec2}.

Equation \eqref{single} describes the phonon-induced resistivity  
for a 1D electron channel with a single pair of Fermi points and agrees with previous 
results \cite{Gurevich1995b,Seelig2005,Dorn2020}.  In particular, 
we obtain a linear dependence $\rho_j\propto T$ for
$T\gg T_{\rm BG}$.  However, for $T\ll T_{\rm BG}$, Eq.~\eqref{single} predicts an exponentially small 
resistivity, $\rho_j\propto e^{-T_{\rm BG}/T}$, since the probability for having
thermal phonons with the energy required for $2k_F$ scattering processes is exponentially small.

\subsection{Two pairs of Fermi points} \label{sec3d}

\begin{figure}
\includegraphics[width=0.45\textwidth]{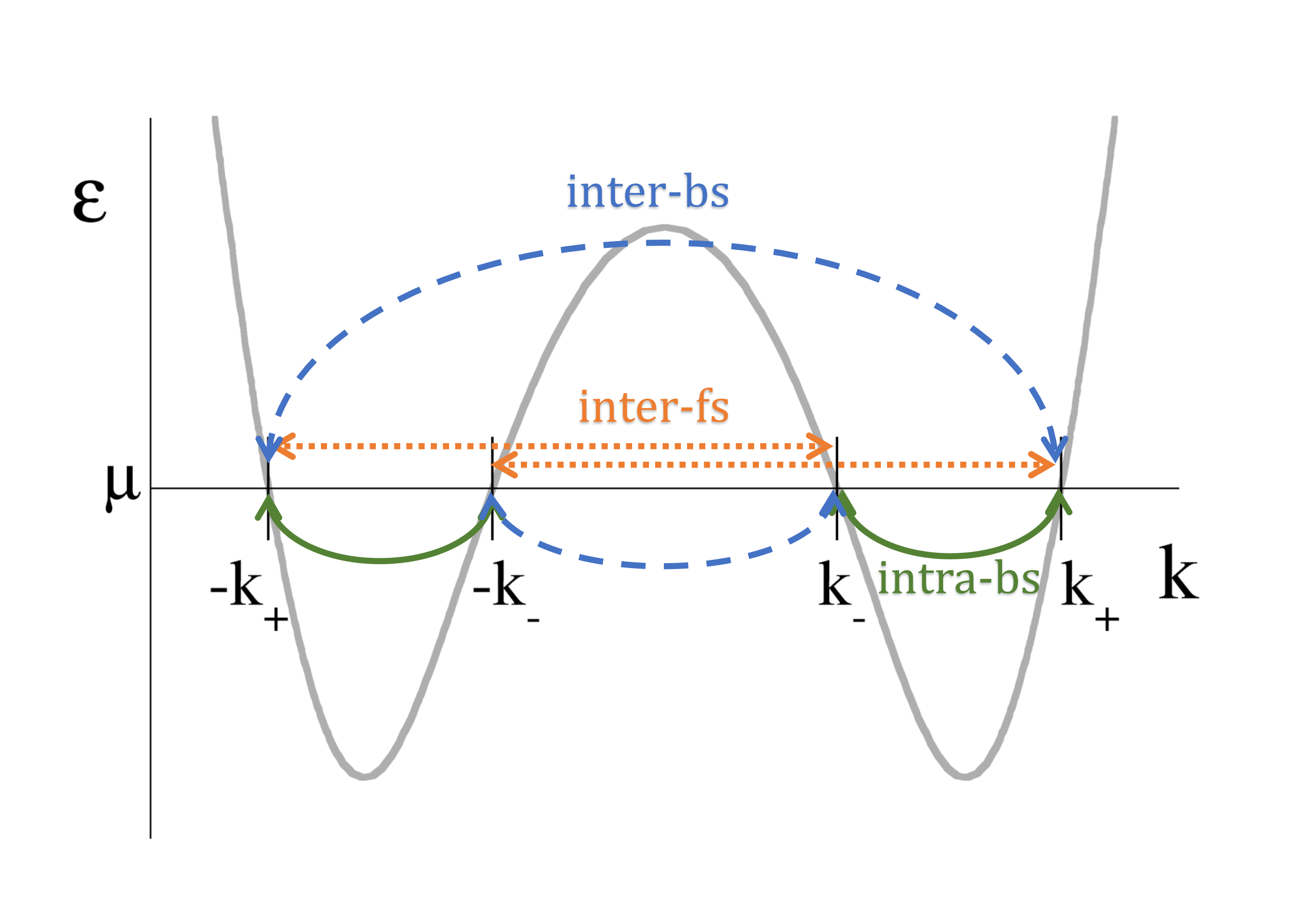}
\caption{Schematic illustration of the different types of scattering processes 
contributing to the resistivity $\rho_j$ for a two-valley subband with two pairs of Fermi points, see Sec.~\ref{sec3d}.}
\label{fig5}
\end{figure}

Next we turn to the resistivity contribution generated by a two-valley band  with the 
Fermi level adjusted to realize two pairs of Fermi points $k= \pm k_{\gamma=\pm}$, with 
Fermi momenta $k_+>k_->0$ and Fermi velocities $v_\gamma=|\partial_k \varepsilon_{k=k_\gamma}|$.
Note that the group velocities for $k\sim k_+$ and $k\sim k_-$ have opposite sign. 
Three different scattering channels are now important, 
see Fig.~\ref{fig5}.  In particular, we distinguish the following processes:
\begin{enumerate}
\item  In analogy to $2k_F$ scattering, see Sec.~\ref{sec3c}, we have
\emph{inter-node backscattering} (``inter-bs'') processes,
where an electron scatters between $k\sim k_{\gamma}$ and $k'\sim - k_{\gamma}$ (with $\gamma=\pm$). 
The momentum exchange $2k_\gamma$ has to be supplied by phonons.
\item For a two-valley band, the dispersion has two local extrema inherited from the 
Weyl nodes at $k=\pm b$.  As a consequence, for appropriate values of the chemical potential, 
\emph{intra-node backscattering} (``intra-bs'') processes become possible, where scattering takes place between
$k\sim sk_{+}$ and $k'\sim sk_{-}$ with $s=\pm$.  
Since the momentum transfer $k_+-k_-$ is typically small against the other relevant momentum transfers,
the contributions due to intra-bs processes are particularly important at low temperatures.
\item Finally, \emph{inter-node forward scattering} (``inter-fs'') processes 
couple states with the same sign of the velocity, i.e., $k\sim sk_{+}$ and $k'\sim -s k_{-}$.
Even though right movers scatter to right movers again, and similarly for left movers,
resistivity contributions arise because of the velocity change for $v_+\ne v_-$.  
We note that forward scattering processes near a single Fermi point are always negligible, see App.~\ref{appC}.
\end{enumerate}

Repeating the analysis of Sec.~\ref{sec3c} for two pairs of Fermi points, see App.~\ref{appC} for details, 
the solution of the Boltzmann equation follows from Eq.~\eqref{CB} with $C\simeq ( v_++v_-)/\pi$ and 
\begin{equation} 
  A \simeq A_{\rm inter-bs}+A_{\rm intra-bs}+A_{\rm inter-fs}   . \label{bulkcoeff}
\end{equation}
The inter-bs contribution is given by, cf.~Eq.~\eqref{afinal},
\begin{equation}
A_{\rm inter-bs} \simeq \frac{4}{\pi} Z v^2 \sum_{\gamma=\pm} k_\gamma \,
{\cal F}\left(T_{\rm inter-bs}^{(\gamma)}/T\right),
\end{equation}
with ${\cal F}(X)$ in Eq.~\eqref{auxfunc} and the Bloch-Gr\"uneisen scales 
$T_{\rm inter-bs}^{(\pm)}=2c_L k_\pm.$
Intra-bs processes imply the contribution
\begin{eqnarray}\nonumber
A_{\rm intra-bs} &\simeq & \frac{1}{\pi} Z v^2 \frac{(v_++v_-)^2}{v_+ v_-}
   (k_+-k_-) \, {\cal I}_{k_+,k_-} \\  \label{aintrabs}
&\times& {\cal F}\left (T_{\rm intra-bs}/T\right)
\end{eqnarray}
with the overlap matrix element \eqref{overlap} and the Bloch-Gr\"uneisen scale
$T_{\rm intra-bs}=c_L (k_+-k_-)$.
Finally, inter-fs contributions are given by
\begin{eqnarray}\nonumber
A_{\rm inter-fs} &\simeq & \frac{1}{\pi} Z v^2 \frac{(v_+- v_-)^2}{v_+ v_-}
   (k_++k_-)\, {\cal I}_{k_+,k_-} \\  
&\times& {\cal F}\left(T_{\rm inter-fs}/T\right) 
\end{eqnarray}
with $T_{\rm inter-fs}=c_L (k_++k_-)$.  We here used ${\cal I}_{k_+,-k_-}={\cal I}_{k_+,k_-}$, 
which holds because the radial spinor eigenfunctions $Y_k(\xi)$ only depend on $|k|$.
\begin{figure*}
\includegraphics[width=5.9cm]{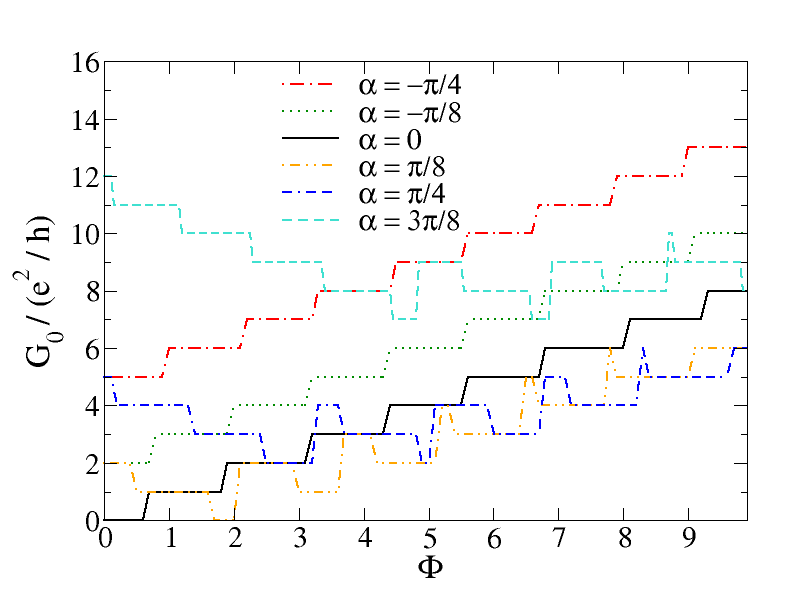}
\includegraphics[width=5.9cm]{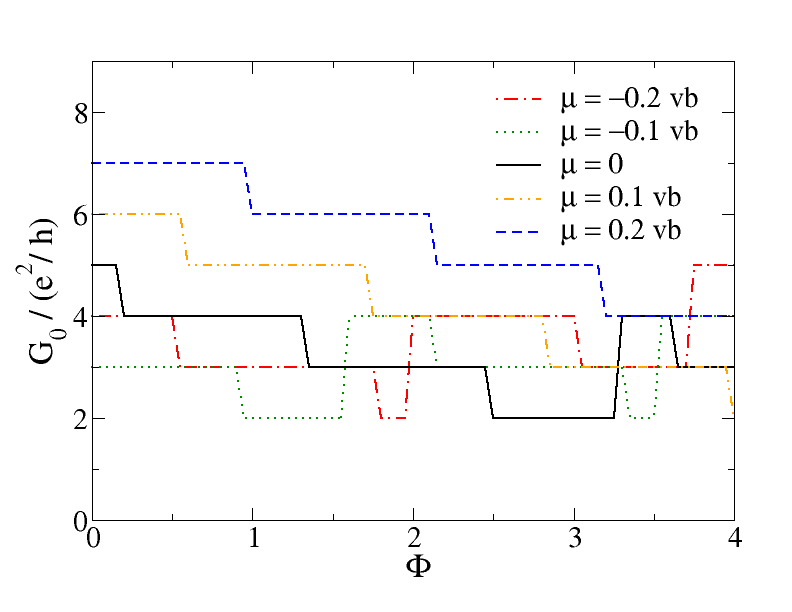}
\includegraphics[width=5.9cm]{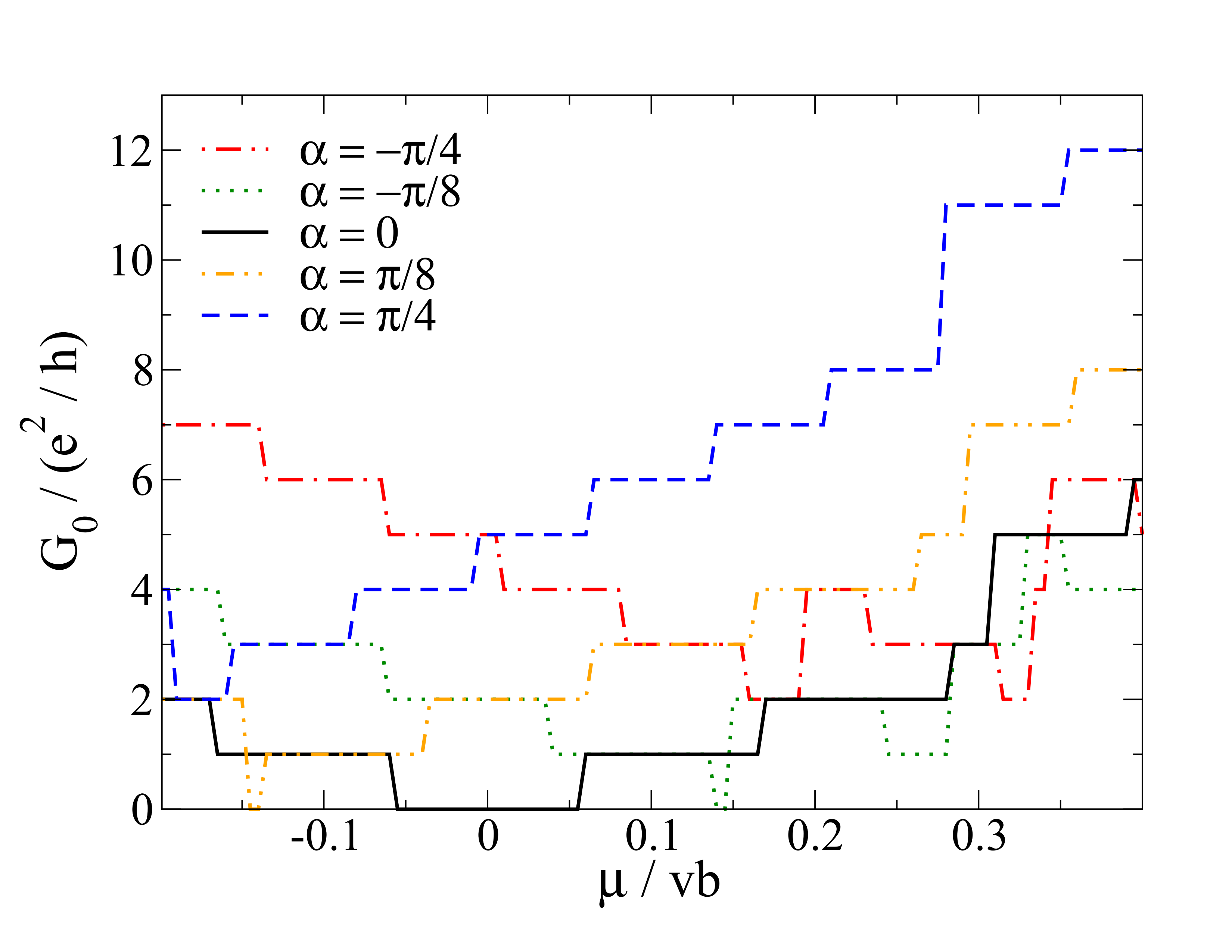}
\caption{Zero-temperature two-terminal conductance $G_0$ of a nanowire with $bR=10$ as obtained from Eq.~\eqref{conductance}. The
left and center panels show the dependence on the magnetic flux parameter $\Phi$ for $\mu=0$ and several values of $\alpha$ (left), and for
$\alpha=\pi/4$ and several values of $\mu$  (center). The right panel shows the dependence on $\mu$
for $\Phi=0$ and several values of $\alpha$.}
\label{fig6}
\end{figure*}

Collecting all terms, the resistivity contribution $\rho_j$ follows as
\begin{equation}\label{bulkcon}
   \rho_j = \rho_{{\rm inter-bs}} + \rho_{\rm intra-bs} + \rho_{\rm inter-fs} .
\end{equation}  
With the reference scale $\rho_0$ in Eq.~\eqref{rho0}, we obtain
\begin{eqnarray}\nonumber
\frac{\rho_{\rm inter-bs}}{\rho_0} &=&  \sum_\gamma \frac{2k_\gamma}{b} \frac{v^2}{(v_++v_-)^2} \, {\cal F}\left(T_{\rm inter-bs}^{(\gamma)}/T\right),\\
\nonumber
\frac{\rho_{\rm intra-bs}}{\rho_0} &=&  \frac{k_+-k_-}{b} {\cal I}_{k_+,k-} \frac{v^2}{2v_+ v_-} \, {\cal F}\left(T_{\rm intra-bs}/T\right),\\
    \frac{\rho_{\rm inter-fs}}{\rho_0} & = & \nonumber
    \frac{k_++k_-}{b} {\cal I}_{k_+,k-} \left(\frac{v_+-v_-}{v_++v_-}\right)^2 \times
    \\ &\times&
    \frac{v^2}{2v_+ v_-} \, {\cal F}\left(T_{\rm inter-fs}/T\right). \label{rhocont}
\end{eqnarray}
From Eq.~\eqref{bulkcon}, the contributions from different backscattering channels simply 
add up and Mathiessen's rule \cite{Landau10} seems to be valid. 
However, Mathiessen's rule is \emph{not} valid for the two different inter-bs processes related to $2k_+$ and $2 k_-$ backscattering,
which cannot be treated separately because of the factor $1/(v_++v_-)^2$ in $\rho_{\rm inter-bs}$.
We stress that in Eq.~\eqref{rhocont}, the quantities $k_\pm$ and $v_\pm$, and thus also
the overlap integral ${\cal I}_{k_+,k_-}$ and the various Bloch-Gr\"uneisen temperatures, depend on the 
specific subband under consideration, in particular on the angular momentum $j$.

In general, the scattering channel with the smallest of the above Bloch-Gr\"uneisen scales
(denoted by $T_{\rm bBG}$) dominates the low-temperature resistivity.
In particular,  $\rho_j\propto T$ for $T\gg T_{\rm bBG}$ while 
$\rho_j\propto e^{-T_{\rm bBG}/T}$ for $T\ll T_{\rm bBG}$.  
In many cases of interest,  $T_{\rm bBG}$ can be well
below the inter-bs scale $T_{\rm BG}$. 
The low-temperature resisitivity is thus dominated by those subbands 
which allow for intra-bs processes.

\section{Transport observables}\label{sec4}

In this section, we describe our results for transport observables.  
In Sec.~\ref{sec4a}, we consider the two-terminal conductance for an ideal WSM nanowire 
in the zero-temperature limit, where phonons are frozen out. 
The conductance is then directly determined by the total number of transport channels 
at the Fermi level. In Sec.~\ref{sec4b}, we present results for the 
phonon-induced resistivity as obtained from the Boltzmann theory in Sec.~\ref{sec3}.

\subsection{Conductance of ideal WSM nanowires}\label{sec4a}

We first consider the two-terminal linear magneto-conductance of a WSM nanowire 
without disorder and in the absence of electron-phonon interactions, 
assuming perfectly adiabatic contacts
between the nanowire and the attached source and drain electrodes.  
This problem can be described by the Landauer-B\"uttiker scattering approach \cite{Nazarov2009}, 
which implies that the two-terminal conductance $G_0$ is given by \cite{Igarashi2017,Kaladzhyan2019,Sukhachov2020}
\begin{equation}\label{conductance}
    G_0(\mu,\Phi,\alpha) = N \frac{e^2}{h},
\end{equation}
where $N=N(\mu,\Phi,\alpha)$ is the number of transport channels at the Fermi level, 
which coincides with the number of positive Fermi momenta. 
The conductance in Eq.~\eqref{conductance} then follows directly from the band structure in Sec.~\ref{sec2}. 
We note that $G_0$ has been studied before for WSM nanowires with boundary conditions corresponding to $\alpha=0$~\cite{Igarashi2017,Kaladzhyan2019,Sukhachov2020}. Our results are consistent with those works
and extend them to arbitrary values of $\alpha$.  

We illustrate  the dependence of $G_0$ on the magnetic field in Fig.~\ref{fig6}, 
both for chemical potential $\mu=0$ and various $\alpha$ (left panel), and for 
$\alpha=\pi/4$ and several values of $\mu$ (center panel). 
The number $N$, and thus $G_0$, jumps in discrete units upon changing $\Phi$.  
The addition (or removal) of one pair of Fermi points to (from) the Fermi surface implies conductance steps of size 
$\Delta G_0=\pm e^2/h$ from Eq.~\eqref{conductance}.  We also see steps with $\Delta G_0=\pm 2 e^2/h$,
where a two-valley band with two pairs of Fermi points is added or removed.

The flux dependence shown in Fig.~\ref{fig6} reveals that conductance steps occur with a
typical spacing of order $\Delta \Phi \approx 1$.  
To rationalize this observation, we recall that the Fermi 
arc dispersion depends on the Aharonov-Bohm phase through the shift $j\to j+\Phi$, 
see Eq.~\eqref{fermiarc}.  Changing $\Phi\to \Phi+1$ shifts the 
sequence of surface subbands by one unit. In a surface-dominated regime, conductance 
variations thus have the (approximate) period $\Delta \Phi\approx 1$.  
Similar features have been experimentally observed in Dirac semimetal wires \cite{Wang2016,Lin2017}.

From the left panel of Fig.~\ref{fig6}, we observe that the boundary angle $\alpha$ 
has a major impact on the conductance. This strong sensitivity of $G_0$ on a boundary parameter is 
consistent with the fact that for the parameters in Fig.~\ref{fig6}, 
we mainly have surface states at the Fermi level. 
In our model, the phenomenological parameter $\alpha$ encodes the surface feature 
of the WSM material. This sensitivity thus indicates that the surface structure of 
the material can strongly influence the conductance. 

The rich band structure exemplified in Fig.~\ref{fig3} also implies that the 
two-terminal conductance is not a monotonic function of the magnetic flux. 
In an infinite WSM, a negative magnetoresistance is expected when ${\bf E} \parallel {\bf B}$, 
as a direct consequence of the chiral nature of the lowest Landau levels. In our cylindrical geometry, 
the spectrum is qualitatively very different from the bulk case, hence one may expect a different behavior.
Indeed, as seen in the left panel of Fig.~\ref{fig6} for $0\le\alpha<\pi/2$, 
the magnetoconductance shows a non-monotonic behavior with a minimum at 
$\Phi\approx \Phi_{\rm min}(\alpha)$, even for the clean case under consideration,
and strongly depends on the surface parameter $\alpha$. This
non-monotonicity of the magnetoresistance is a manifestation of the predominance of the 
surface over the bulk transport in this geometry.
Interestingly,
the value of $\Phi_{\rm min}$ can be determined by an approximate fit of 
$G_0(\Phi)$ to a third-order polynomial function. For the conductance curves shown 
in the left panel of Fig.~\ref{fig6}, we observe that $\Phi_{\rm min}$ is linked 
to the boundary angle by the empirical relation $\alpha\simeq 0.28\Phi_{\rm min}-0.01\Phi_{\rm min}^2$.
By determining the position of the magnetoconductance minimum, one can thus infer 
information about $\alpha$ from transport measurements, at least in the parameter regime under study here.

In analogy to the stepwise dependence on the flux, we 
also find conductance steps when varying $\mu$ at fixed magnetic flux, 
as shown in the right panel of Fig.~\ref{fig6} for several values of $\alpha$.
For $\alpha=0$, this parameter region was identified in Ref.~\cite{Kaladzhyan2019}, via the conductance steps, as the 
regime in which surface states dominate transport. Our results confirm this scenario.
At the same time, we observe that a finite value of the boundary angle $\alpha$ can dramatically change  
the low-temperature transport properties. In fact, only for special values of $\alpha$, we obtain
insulating behavior at zero magnetic field and $T \ll v/R$.  For generic $\alpha$,  the two-terminal conductance is 
finite and can even become large. This observation again highlights the importance of non-universal surface physics in this geometry.

Finally, we note that even though we have a finite two-terminal conductance $G_0$, the 
local resistivity $\rho$  vanishes in the absence of phonon-induced (or other) backscattering 
processes. 

\begin{figure*}[h]
\includegraphics[width=8cm]{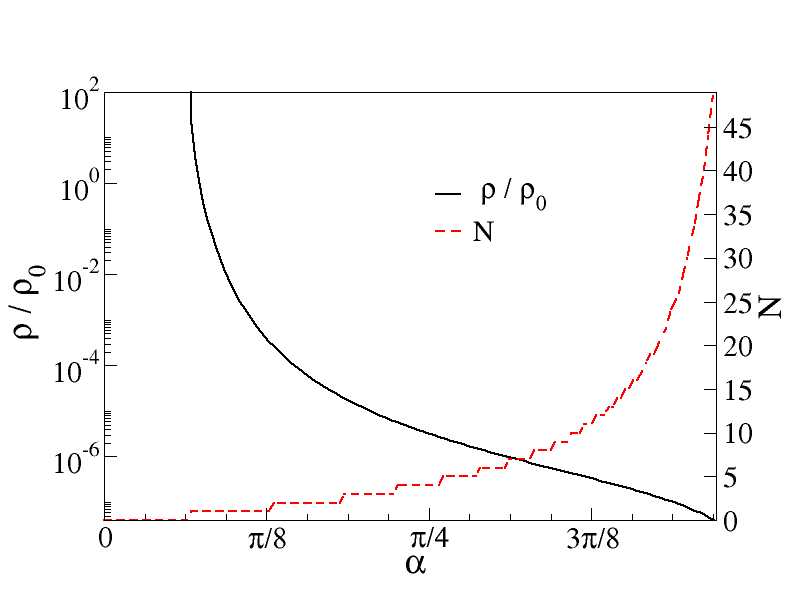}
\includegraphics[width=8cm]{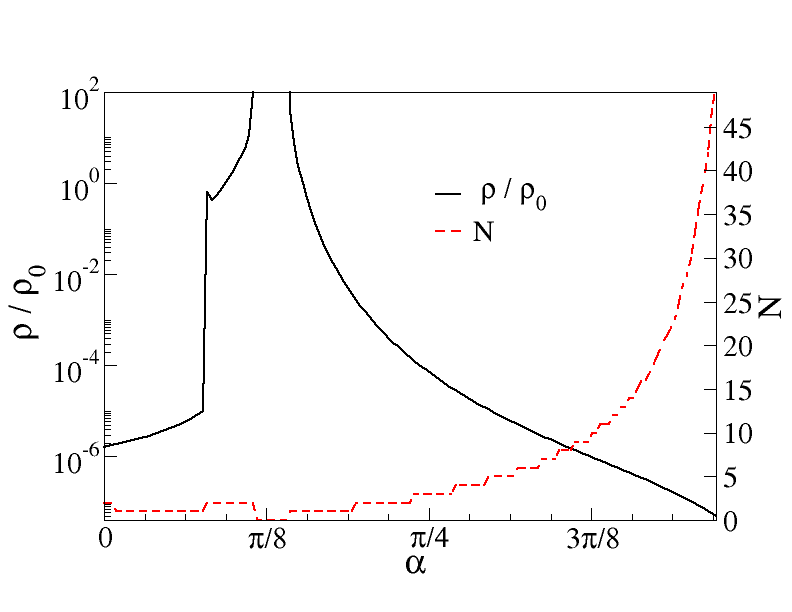}
\caption{
Resistivity $\rho$ (in units of $\rho_0$) vs boundary angle $\alpha$ for $T=0.1c_Lb$, 
$\mu=0$, $bR=10$, and $c_L=0.01v$, with magnetic flux parameter  $\Phi=1/2$ (left panel) 
and $\Phi=2$ (right panel). We use logarithmic scales for $\rho/\rho_0$ (solid black curves). 
The number of Fermi points $N$ is shown by red dashed curves. The divergence at small values 
of $\alpha$ in the left panel and around $\alpha=\pi/8$ in the right panel is due to the fact that 
for these values there are no available bands at the Fermi level. 
}
\label{fig7}
\end{figure*}

\begin{figure*}
\includegraphics[width=8cm]{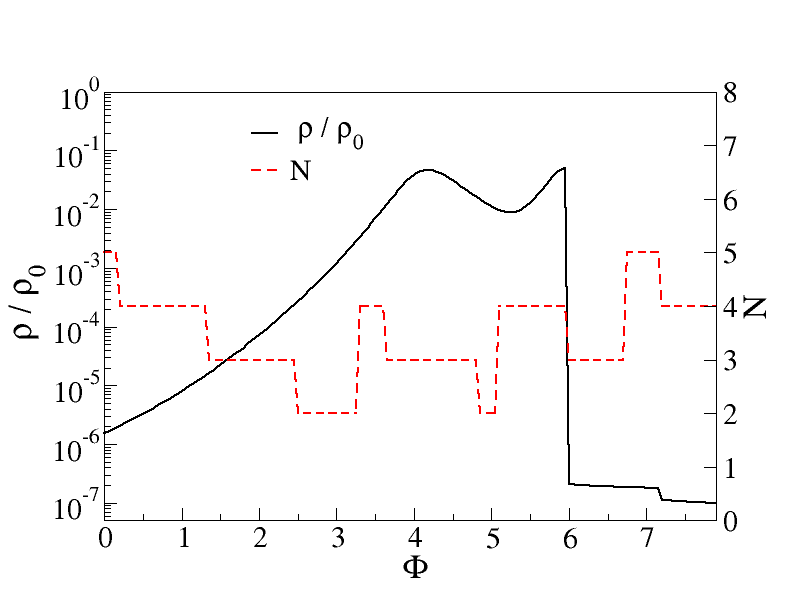}
\includegraphics[width=8cm]{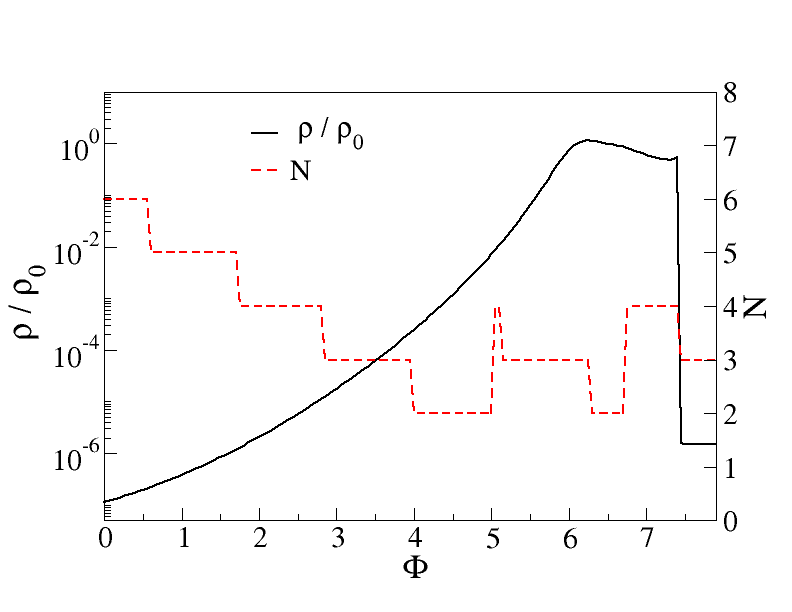}
\caption{Resistivity $\rho/\rho_0$ vs magnetic flux parameter $\Phi$ 
for $\alpha=\pi/4$ and $T=0.1 c_L b$, with $\mu =0$ (left panel) and $\mu=0.1vb$
(right panel).  All other parameters and conventions are as in Fig.~\ref{fig7}.
}
\label{fig8}
\end{figure*}

\begin{figure*}
\includegraphics[width=8cm]{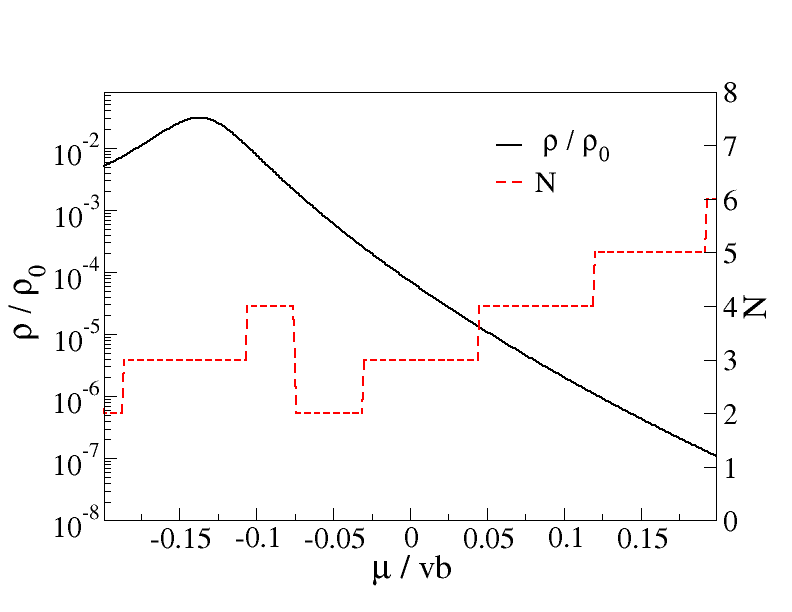}
\includegraphics[width=8cm]{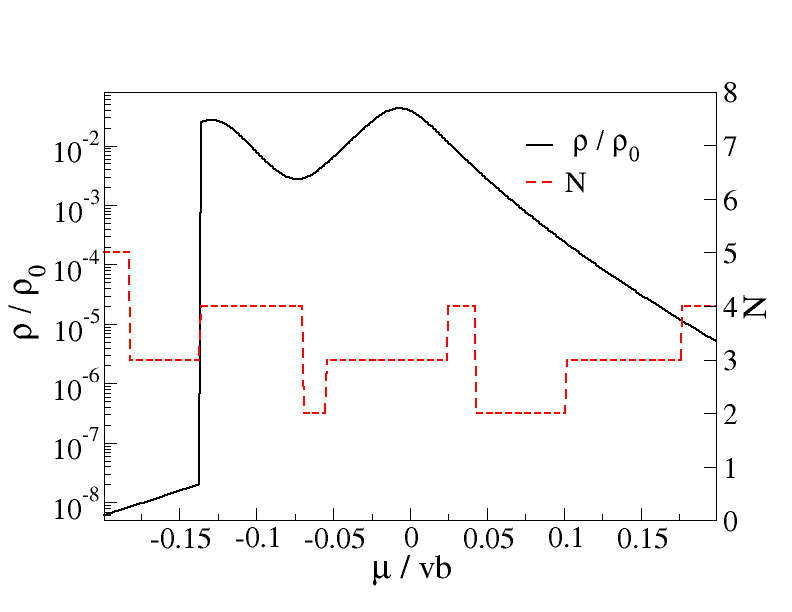}
\caption{Resistivity $\rho/\rho_0$ vs  $\mu$ for $\alpha=\pi/4$ and $T=0.1c_Lb$, with $\Phi =2$ 
(left panel) and $\Phi=4$ (right panel). All other parameters and conventions are as in Fig.~\ref{fig7}.  }
\label{fig9}
\end{figure*}

\begin{figure}[h]
\includegraphics[width=0.45\textwidth]{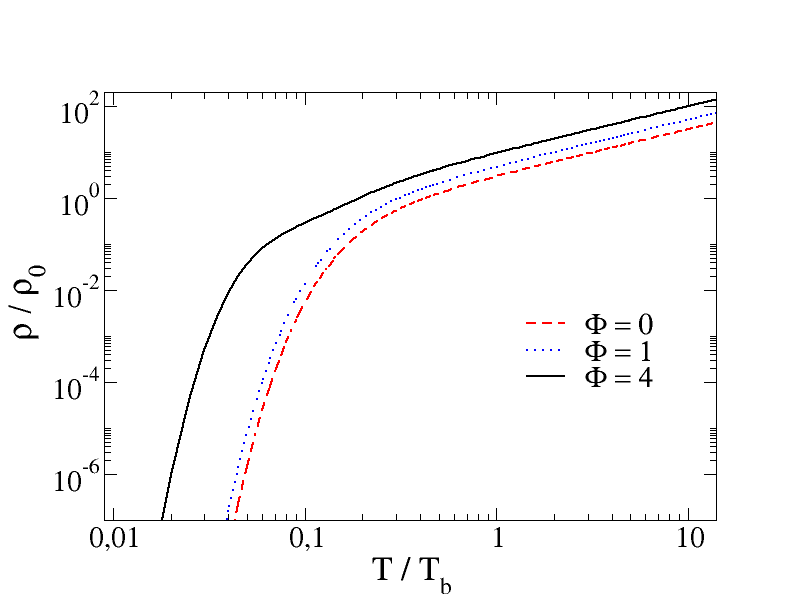}
\caption{Resistivity $\rho/\rho_0$ 
vs temperature $T$ (in units of $T_b=2c_L b)$ for $\alpha=\pi/4$, $\mu=0$, 
and several values of  $\Phi$. Note the double-logarithmic scales.  
All other parameters and conventions are as in Fig.~\ref{fig7}. }
\label{fig10}
\end{figure}

\subsection{Phonon-induced resistivity}
\label{sec4b}

We here discuss our results for the phonon-induced longitudinal magnetoresistivity \eqref{rho1} 
obtained in Sec.~\ref{sec3} using the semiclassical Boltzmann approach. 
We start by illustrating the $\alpha$-dependence 
of $\rho$ for fixed chemical potential $\mu=0$ and temperature $T=0.1 c_Lb$ in Fig.~\ref{fig7}.
While it is not possible to experimentally change the boundary angle $\alpha$ in a given device, 
Fig.~\ref{fig7} shows that the resistivity strongly depends on  $\alpha$.
Typically, with increasing $\alpha$, 1D subbands with different $j$ fall 
below the Fermi level one by one.
As a consequence, the number $N$ increases and the resistivity tends to 
become smaller according to Eq.~\eqref{rho1}. Once a new subband becomes just accessible, 
the corresponding resistivity contribution 
will become very 
large because of the smallness of the Fermi velocity and of the Fermi momentum in this
limit. From Eq.~\eqref{rho1}, we see that such a contribution makes little difference 
as long as other subbands with finite $\rho_j$ are present.  The dependence of $\rho$ 
on $\alpha$ (or other parameters) thus remains smooth even when $N$ changes, with an 
important exception discussed below.

For the parameters corresponding to the left panel in Fig.~\ref{fig7}, where $\Phi=1/2$, only  $j>0$ 
bands with a single pair of Fermi points contribute.  The expected smooth decrease of $\rho(\alpha)$ 
with increasing $\alpha$ is observed. In particular, for small $\alpha$, there are no bands at the Fermi level
and thus $\rho\to \infty$. On the other hand, for $\alpha\to \pi/2$, the resistivity becomes 
extremely small since $N$ increases to very large values. 
The right panel of Fig.~\ref{fig7} shows that for $\Phi=2$, the $\alpha$-dependence of the
resistitivty is more complex. In a finite window 
around $\alpha\approx \pi/8$, $N$ vanishes and $\rho\to\infty$. For $\alpha\agt \pi/8$, 
only $j>0$ bands with a single pair of Fermi points are present, and $\rho(\alpha)$ 
shows a smooth decrease again. For $\alpha\alt \pi/8$,  we have contributions from subbands with $j=-1/2$ and $j=-3/2$.  
At a critical value of $\alpha$  slightly above $\pi/16$, a transition from one to two pairs 
of Fermi points takes place \emph{within} the two-valley subband with $j=-1/2$.  As detailed below 
and in App.~\ref{appD}, such a transition causes an abrupt and very large resistivity increase as 
 seen in Fig.~\ref{fig7}. 
This prominent feature arises because only for cases with more than one pair of Fermi points, 
intra-node backscattering processes become possible, see Sec.~\ref{sec3d}. Such processes 
dominate the resistivity at low temperatures.
  
Next, Fig.~\ref{fig8} shows the magnetic field dependence of the resistivity.  
Let us first discuss the case $\mu=0$ (left panel). We again see that $\rho(\Phi)$ 
is a smooth curve except for an abrupt resistivity drop near $\Phi\approx 6$. Recalling 
the logarithmic scales, the resistivity increase is very steep for small $\Phi$.  
Again, the jump-like behavior at $\Phi\approx 6$ takes place at the transition point from two 
to one pairs of Fermi points within the two-valley subband with $j=-1/2$. 
For large $\Phi$, we observe that $\rho(\Phi)$ also shows variations governed by the
Aharonov-Bohm scale $\Delta\Phi\sim 1$, see Sec.~\ref{sec4a}.   
For $\mu=0.1vb$ (right panel in Fig.~\ref{fig8}), we find similar features.

We now turn to Fig.~\ref{fig9}, which shows the $\mu$-dependence of $\rho$. 
While for $\Phi=2$ (left panel), no abrupt resistivity changes occur in the shown chemical potential range, such  
behavior is found for $\Phi=4$ (right panel) near $\mu=\mu_c\simeq -0.136 vb$.  
We can trace this resistivity change to the two-valley subband with $j=-1/2$.
For $\mu<\mu_c$, this band contributes a single pair of Fermi points.
For $\mu>\mu_c$, on the other hand, we get two pairs of Fermi points.
At the transition, $\mu\simeq \mu_c$, the resistivity exhibits a sharp increase.  
We discuss this mechanism in some detail in App.~\ref{appD} for a simple toy model dispersion. 
For $\mu\to \mu_c$ from above, the Bloch-Gr\"uneisen temperature for intra-bs processes
sets the relevant scale, $T_{\rm bBG}= T_{\rm intra-bs}=c_L(k_+-k_-)$, see Sec.~\ref{sec3d}. 
When approaching the transition from the other side, however, only inter-bs processes can take place,
with $T_{\rm BG}=2c_L k_+$. As a consequence, the resistivity is much larger for $\mu>\mu_c$.
We note that the linearized band structure used in Sec.~\ref{sec3d} is not applicable for $\mu\to \mu_c$.  
However, while the precise $\mu$-dependence of $\rho$ is expected to be continuous when going beyond the 
linearized band structure, the large low-temperature resistivity changes predicted here
should be robust.  
 
Finally, we briefly turn to the temperature dependence of $\rho$, which is shown for $\mu=0$ 
and different $\Phi$ in Fig.~\ref{fig10}. For $T\gg T_b=2c_L b$, we find a universal 
$\rho\propto T$ dependence, but for $T\to 0$, the resistivity becomes exponentially small 
since all phonon backscattering mechanisms are frozen out in that limit. 

\section{Conclusions} \label{sec5}

In this work, we have discussed magnetotransport in a cylindrical WSM nanowire.  
Our analysis includes the effects of a magnetic flux threading the wire (via the Aharonov-Bohm flux $\Phi$)
and the consequences of a finite curvature of the Fermi arc 
(via the boundary angle $\alpha$).  
We have presented detailed results for the band structure, 
in particular how the dispersion of Fermi arc states depends on $\Phi$ and $\alpha$.
The magnetic flux is here 
effectively captured by the replacement $j\to j+\Phi$, where $j$ is 
the half-integer angular momentum of the Fermi arc state.
Importantly, we have taken into account the electron-phonon interaction
via deformation potential. We have focused on phonon modes with 
zero angular momentum, since for nanowires 
deposited on a substrate, phonon modes with finite angular momentum 
are expected to be gapped.

Our analysis shows that the phonon-induced resistivity contains rich information about the
underlying physics of the WSM material. 
The resistivity strongly depends on the boundary
angle $\alpha$ and on the magnetic flux parameter $\Phi$.  
We find that large and abrupt changes of the resistivity arise because of the mexican hat
shape of the dispersion for two-valley subbands, where a change of the chemical potential can induce a transition between 
one vs two pairs of Fermi points. Since in the case of two pairs of Fermi points intra-node backscattering processes
with small momentum transfer are possible, a much larger low-temperature resistivity is obtained than for
the case with a single pair of Fermi points, where such processes are not available.

Comparing our results for WSM nanowires to the case of conventional quantum wires  
\cite{Voit1987,Bockelmann1990,Shik1993,Mickevicius1993,Gurevich1995,Gurevich1995b,Seelig2005,Yurkevich2013},
we find a noteworthy difference.  
Even though it is difficult to quantify the impact of chiral anomaly
on the phonon-induced magnetoresistivity in this finite-size wire geometry, 
the observed strong sensitivity of the resistivity on
a boundary condition parameter is in marked contrast to the conventional setting and can be rationalized by 
the crucial role of Fermi-arc surface states.

Our work also points to several topics of interest for future studies:
(i) For freely suspended WSM nanowires, phonon modes with finite angular momentum have to be included.
In particular, flexural modes with $l=\pm 1$ will be the energetically lowest modes
\cite{Landau7}. One then has to account for scattering 
processes connecting subbands with different angular momenta.  
(ii) Similarly, at higher energy scales and/or very large nanowire radius, 
the restriction to a single radial band for given angular momentum $j$ 
has to be lifted even when keeping only $l=0$ phonon modes.    One may then
encounter more than two pairs of Fermi points at fixed angular momentum $j$, and many additional scattering
processes beyond those considered in Sec.~\ref{sec3} become possible. 
(iii) The above two points are important also for the proper description of nonequilibrium
transport beyond the linear response regime considered here. 
(iv) In the present work, we have studied type-I WSM materials.  In type-II WSM materials, one 
has (over-)tilted Dirac-Weyl cones with interesting analogies to black hole physics \cite{Kedem2020}.
In such a setting, phonons may give spectacular effects, cf.~Ref.~\cite{Gomez2021}.
(v) At very low temperatures, disorder effects will dominate the resistivity in real samples.  
While the zero-field resistivity of disordered WSM nanowires (without phonon effects) 
has been studied in Ref.~\cite{Gorbar2016}, the magnetoresistivity has not been analyzed 
in a systematic way so far.
(vi) In this work, we have neglected the Zeeman effect due to the magnetic field.  
While one expects such effects to be subleading \cite{Ramshaw2018}, for a precise comparison 
to future experimental results, it may be necessary to include them into the theoretical description.
(vii) An interesting generalization of our work could study WSM materials with more than two Weyl nodes. 
For instance, if the material enjoys time-reversal symmetry at zero magnetic field, there 
will be at least four Weyl nodes. In the presence of phonons and in a magnetic field, 
one then expects a multitude of possible scattering processes.
(viii) Our theory assumes angular momentum conservation. Indeed, we consider a cylindrical wire geometry, 
where the magnetic field is aligned both with the wire axis and with the direction 
of the separation between Weyl nodes in reciprocal space. A weak violation of these conditions 
could be handled by perturbation theory, but for stronger deviations, one has to resort to 
a generalization of our theory and a corresponding numerical study. 
(ix) Finally, apart from the real magnetic field, it may be of interest to study the
consequences of pseudo-magnetic fields generated by straining the sample~\cite{Ilan2020}. 

To conclude, we hope that our paper will stimulate future work along these or other directions.

\begin{acknowledgements}
We thank M.\ Breitkreiz, P.\ W.\ Brouwer, A.\ Kundu, and R.\ G.\ Pereira for discussions.
We acknowledge support by the Deutsche Forschungsgemeinschaft (DFG) 
under Grant No.~EG 96/12-1, under Projektnr.~277101999 - TRR 183 (CRC project A02), 
and under Germany's Excellence Strategy - Cluster of Excellence Matter and Light for Quantum Computing (ML4Q) EXC 2004/1 - 390534769.\\
\end{acknowledgements}

\appendix

\section{Band structure for $\alpha=\pi/2$} \label{appA}

\begin{figure*}
\includegraphics[width=8cm]{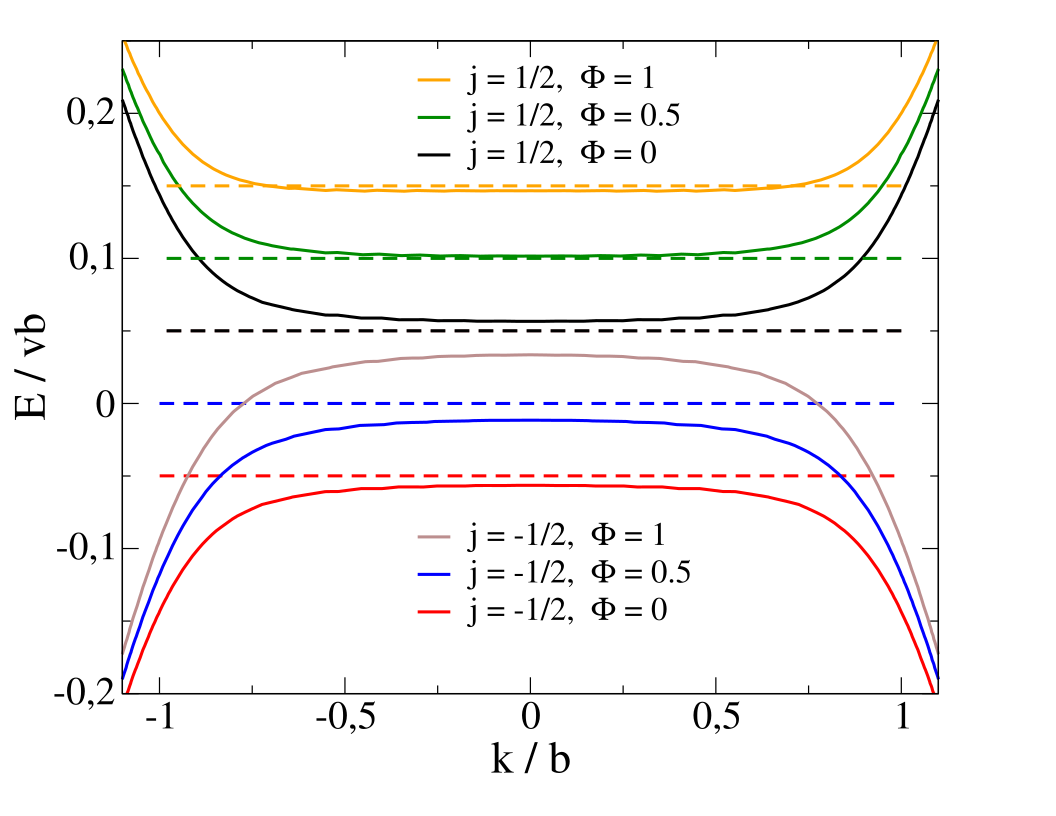}
\includegraphics[width=8cm]{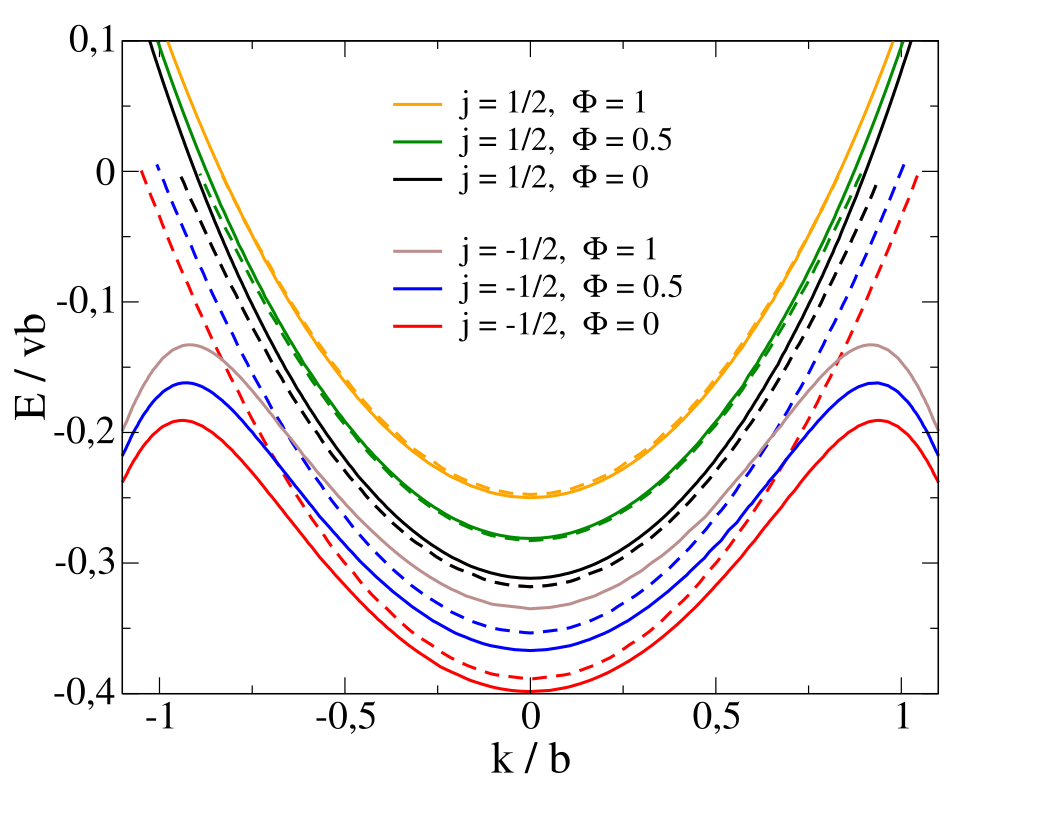}
\caption{Dispersion relation for Fermi arc states with $j=\pm 1/2$ for several values of $\Phi$.
Results are shown for $bR=10$ with $\alpha=0$ (left panel) and $\alpha=\pi/4$ (right panel).
The solid curves were obtained by numerical solution of Eq.~\eqref{bc2}. The dashed curves follow from
the approximate analytical dispersion relation \eqref{fermiarc} and terminate
according to Eq.~\eqref{cond1}. 
Note that the analytical (but not the numerical) results for $(j,\Phi)=(-1/2,1)$ and $(1/2,0)$ coincide.
 }
\label{fig11}
\end{figure*}

We here summarize the exact band structure for the special value $\alpha=\pi/2$, 
where the boundary condition \eqref{bc2} simplifies to $Y_-(\Phi)=0$.
Then, for $j>0$, Eq.~\eqref{psiplus} gives a solution either for ${\cal E}_-=0$ 
(band index $p=0$) or from the zeros of the confluent hypergeometric functions 
(with respect to the first argument), 
$a=a_{j,p}$ with  
 $p=1,2,\ldots$, solving $M(a,j+3/2;\Phi)=0$. 
 Using Eqs.~\eqref{epm} and \eqref{adef}, the dispersion relations of the respective subbands follow as 
\begin{equation}\label{bands}
E_{k,j>0,0}=  m_k, \quad E_{k,j,\pm p} = \pm \sqrt{2 C_{j,p}  (v/l_B)^2 + m^2_k},
\end{equation}
with $C_{j>0,p} = j+1/2 - a_{j,p}$; the zeros $a_{j,p}$ are all negative numbers.
The radial eigenfunctions \eqref{PHidef} for $p=0$ are given by
\begin{equation} \label{eigspecial1}
Y_{k,j>0,0}(\xi) \propto \xi^{\frac12(j-\frac12)} e^{\xi/2} \begin{pmatrix}
1 \\ 0 \end{pmatrix}.
\end{equation}
The associated probability density increases with $\xi$ and has a maximum at the surface, i.e., for $\xi=\Phi$.
The dispersion relation $\varepsilon_k=m_k$ for the degenerate $p=0$ subbands with $j>0$ agrees 
with the Fermi arc dispersion for $\alpha=\pi/2$ in Eq.~\eqref{fermiarc}.
On the other hand, for $j<0$, the $p=0$ band does not exist at finite $R$, and
all $p\ne 0$ bands occur in pairs as follows from Eq.~\eqref{bands} with $C_{j<0,p}=-a_{-j-1,p}$.
The band structure for $\alpha=\pi/2$ is illustrated in  Fig.~\ref{fig1} for few different 
values of $\Phi$.

Let us further discuss Eq.~\eqref{bands} in two limiting cases. 

\noindent (i) For $\Phi\to 0$, Eq.~\eqref{bands} reduces to
\begin{equation}\label{bands2}
E_{k,j>0,0}=  m_k, \quad E_{k,j,\pm p} = \pm \sqrt{(v z_{j,p} /R)^2 + m^2_k},
\end{equation}
where $z_{j,p}>0$ is the $p$th zero of the Bessel function $J_{j+1/2}(z)$.  
The $p\ne 0$ bands correspond to bulk states, 
which involve the finite-size quantization energy scale $v/R$. 
The states in the $p=0$ bands have radial eigenfunctions
$Y_{k,j>0,0}(r) \propto r^{j-\frac12} \begin{pmatrix}
1 \\ 0
\end{pmatrix}$ and correspond to degenerate Fermi arc surface states.
From Eq.~\eqref{bands},  we observe that the dispersion relation of the $p=0$ subbands
is not affected by the magnetic field, although the states are.  (The $j=1/2$ state is obviously not localized at the surface. 
However, taking the limit of large $j$ and large $R$ at fixed ratio $j/R$, the corresponding states represent \emph{bona fide} surface states.)  

\noindent (ii) For very large but finite $\Phi$, the zeros of the confluent hypergeometric functions 
approach negative integer values, $a_{j,p}\to -(p-1)$.  
As a consequence, we recover the bulk Landau level spectrum \eqref{bulk}.  
The Fermi arc states with $p=0$ and $j>0$ exist for any finite $\Phi$ but disappear
in the limit of infinite radius.  
From Eq.~\eqref{bands}, we also observe that the dispersion of the bulk states $\Psi_{k,j<0,\pm 1}$  
approaches $\pm |m_k|$, with an avoided crossing at $k=\pm b$.  
The latter is formally due to the fact that $a=0$  is never a solution of $M(a,-j+\frac12;\Phi)=0$.
In the limit $R\to \infty$, the gap closes. The branch with $E=-m_k$ reduces to the usual 
bulk zero mode, see Eq.~\eqref{bulk}, and the branch $E=m_k$ disappears. \\

\section{On Fermi arc surface states}\label{appB}

In this Appendix, we construct approximate surface state solutions 
and compare the analytical result for their dispersion relation 
with the band structure obtained numerically from Eq.~\eqref{bc2}.

Starting from the radial Dirac-Weyl equation for the spinor $Y(r)$ (the indices $k,j$ are understood),
\begin{equation}
    \begin{pmatrix}
    -\frac{1}{v}(E-m_k) &\partial_r + \frac{j+\frac12}{r} + \frac{r}{l^2_B}  \\
- \partial_r + \frac{j-\frac12}{r} +\frac{r}{2l_B^2} &  - \frac{1}{v}(E+m_k)
 \end{pmatrix} Y(r) =0 ,
 \label{radialHam}
\end{equation}
we first write the radial coordinate as $r=R+x$ with $-R<x<0$.  
We search for solutions localized at the surface, with main weight at $|x|\ll R$ 
and decaying for increasing $|x|$. 
Expanding Eq.~\eqref{radialHam} to lowest non-trivial order in $|x|/R\ll 1$ and writing
$Y(r) = e^{(x-R)^2/4R^2}\chi(x)$, we arrive at 
 \begin{equation}\label{approx1}
 \begin{pmatrix}
    -\frac{1}{v}(E-m_k) &\partial_x  + \frac{j+\Phi}{R} - \frac{j -\Phi}{R^2} x \\
- \partial_x + \frac{j+\Phi}{R} - \frac{j - \Phi}{R^2}x &  - \frac{1}{v}(E+m_k)
 \end{pmatrix} \chi(x) =0 .
\end{equation}
This equation can be solved exactly, but we here
consider a simpler approximate solution. 
We neglect the term $\propto x$ in Eq.~\eqref{approx1}, so that 
$\chi(x)\propto e^{\kappa x} \chi(0)$ is a solution, 
with the inverse decay length $\kappa$ given by Eq.~\eqref{kappadef}.
The consistency of the approximation requires $\kappa R\gg 1$.
Imposing the boundary condition \eqref{bc2} on the eigenstate $\chi (0)$, we arrive 
at the dispersion relation \eqref{fermiarc} with the condition \eqref{cond1}.  
To estimate the neglected term $\propto x$ in Eq.~\eqref{approx1},
we put $|x|\sim 1/\kappa$. We then require $|j + \Phi|/R \gg |(j - \Phi) x|/R^2$, 
which in turn implies the condition \eqref{cond2}.

We next compare the approximate dispersion relation 
Eq.~\eqref{fermiarc} to the numerically exact band structure.  In
Fig.~\ref{fig11}, we show the dispersion of Fermi arc states with $j=\pm 1/2$ for $bR=10$ and several values of
$\Phi$ and $\alpha$.  We find a fair agreement between numerical and analytical results. 
In accordance with Eq.~\eqref{cond2}, the deviations are more pronounced for $j<0$ 
and $\Phi\ne 0$, but  even for $j=-\Phi=-1/2$, Eq.~\eqref{fermiarc} provides a rather good approximation.
Since the penetration length $\kappa^{-1}$ becomes very large near the arc ends, 
the analytical expression in Eq.~\eqref{fermiarc} --- 
which assumes $\kappa R\gg 1$ --- becomes less accurate in these limits, in accordance with Fig.~\ref{fig11}.

\section{Solution of the Boltzmann equation}\label{appC}

We present here the derivation of Eqs.~\eqref{afinal} and \eqref{bulkcoeff} 
for one and two pairs of Fermi points, respectively. Following Ref.~\cite{Levchenko2020},
we begin by rewriting the coefficient $A$ in Eq.~\eqref{CB} as
\begin{eqnarray}\label{aaa}
A &=& \frac{1}{2T} \int d\varepsilon d\varepsilon' \int_0^\infty d\omega \, F(\varepsilon,\varepsilon',\omega) \times
\\ &\times& \nonumber
\frac{\omega n_{F}(\varepsilon) n_{F}(\varepsilon') }
{\left|e^{-\beta (\varepsilon-\mu)}-e^{-\beta(\varepsilon'-\mu) }\right|}
\sum_{\nu=\pm} \delta(\varepsilon-\varepsilon' - \nu \omega)
\end{eqnarray}
with the auxiliary function
\begin{eqnarray}  \nonumber
F(\varepsilon,\varepsilon',\omega)&  =& \frac{1}{\omega} 
\int \frac{dk}{2\pi}\frac{dk'}{2\pi}  W(k',k)\left( v_{k'} - v_{k}  \right)^2 \times \\ &\times& \label{fdef}
\delta(\varepsilon-\varepsilon_k) \delta(\varepsilon' -\varepsilon_{k'}) \delta(\omega - \omega_{k-k'}).
\end{eqnarray}
At low temperatures, the momentum integrations in Eq.~\eqref{fdef} 
can be restricted to the vicinity of the Fermi points.

Let us first consider the case of a single pair of Fermi momenta, see Sec.~\ref{sec3c}.
Writing $k=sk_F+\tilde k$ and $k'=s'k_F+\tilde k'$ with $s,s'=\pm$ and  $|\tilde k|,|\tilde k'|\ll k_F$, 
we first linearize the dispersion relation, $\varepsilon_{\pm k_F+\tilde k}-\mu\simeq \pm v_F\tilde k$.  
We then have backscattering contributions to Eq.~\eqref{fdef} when $k$ and $k'$ are near  
opposite Fermi points ($s=-s'$), and forward scattering contributions 
when $k$ and $k'$ are near the same Fermi point ($s=s'$).  
The forward scattering terms are strongly suppressed by the 
factor $(v_{k'}-v_{k})^2\propto (\tilde k-\tilde k')^2$ in Eq.~\eqref{fdef}, 
and they are always neglected in what follows.  
With $v_k\simeq sv_F$, the backscattering contributions follow by approximating $W(k,k')\simeq
W(k_F,-k_F)=W(-k_F,k_F) \equiv W_{\rm bs}$. 
Since the $k$-dependence of the radial eigenfunctions $Y_k(\xi)$ 
arises only through $m_k$, which is an even function of $k$, we have
${\cal I}_{k,-k}={\cal I}_{k,k}$, and the normalization in Eq.~\eqref{overlap} 
implies ${\cal I}_{k,k}=1$. Thus,
with $W_{\rm bs}=4\pi Z v^2 k_F$ from Eq.~\eqref{Wdef}, we obtain 
\begin{equation}\label{Fappr}
F(\varepsilon,\varepsilon',\omega)\simeq  \frac{4 Z v^2}{\pi c_L} \delta(\omega- 2c_L k_F).
\end{equation}
Using the auxiliary relation \cite{Levchenko2020}
\begin{eqnarray}\nonumber
  &&  \int d\varepsilon d\varepsilon' \frac{n_{F}(\varepsilon) n_{F}(\varepsilon') }
{\left|e^{-\beta (\varepsilon-\mu)}-e^{-\beta(\varepsilon'-\mu) }\right|} \times \\ && \times
\sum_{\nu=\pm} \delta(\varepsilon-\varepsilon' - \nu \omega)= \frac{\omega}{2\sinh^2(\beta\omega/2)}
\end{eqnarray}
in Eq.~\eqref{aaa}, we finally arrive at Eq.~\eqref{afinal}. The
above approximations also imply $C\simeq v_F/\pi$ from Eq.~\eqref{CB}.

Next we turn to a two-valley band with the Fermi level adjusted to allow for 
two pairs of Fermi momenta at $k=\pm k_\gamma$ with $\gamma=\pm$, see Sec.~\ref{sec3d} and Fig.~\ref{fig5}.  
The symmetry $\varepsilon_k=\varepsilon_{-k}$ then implies that the group velocity at $k\sim sk_\gamma$ 
is given by $v_{s,\gamma} = s\gamma v_\gamma$ (where $s=\pm$), with the positive Fermi 
velocities $v_+$ and $v_-$.   Linearizing the dispersion relation for 
$k\approx s k_{\gamma}$, contributions to Eq.~\eqref{fdef} from the three
 types of scattering processes illustrated in Fig.~\ref{fig5} arise.  We find
\begin{equation}
    F(\varepsilon,\varepsilon',\omega) \simeq F_{\rm inter-bs}+ F_{\rm intra-bs} + F_{\rm inter-fs} ,
\end{equation}
where, in analogy to the  $2k_F$ backscattering result \eqref{Fappr}, inter-node backscattering 
processes give
\begin{equation}
    F_{\rm inter-bs}\simeq \frac{4 Z v^2}{\pi c_L} \sum_{\gamma=\pm} \delta(\omega- 2c_L k_\gamma).
\end{equation}
Intra-node backscattering processes produce the term
\begin{equation}
    F_{\rm intra-bs} \simeq \frac{2Z v^2}{\pi c_L} \frac{(v_++v_-)^2}{v_+ v_-}
    {\cal I}_{k_+,k_-}  \delta\left(\omega- 2c_L|k_+-k_-|\right),
\end{equation}
with ${\cal I}_{k,k'}$  in Eq.~\eqref{overlap}, and
inter-node forward scattering contributions give 
\begin{equation}
    F_{\rm inter-fs}\simeq \frac{2Z v^2}{\pi c_L} \frac{(v_+- v_-)^2}{v_+ v_-}
    {\cal I}_{k_+,k_-}  \delta\left(\omega- 2c_L|k_++k_-|\right).
\end{equation}
Inserting the above results into Eq.~\eqref{aaa}, we arrive at Eq.~\eqref{bulkcoeff}. \\

\section{Abrupt resistivity changes}\label{appD}

\begin{figure}[t]
\includegraphics[width=0.5\textwidth]{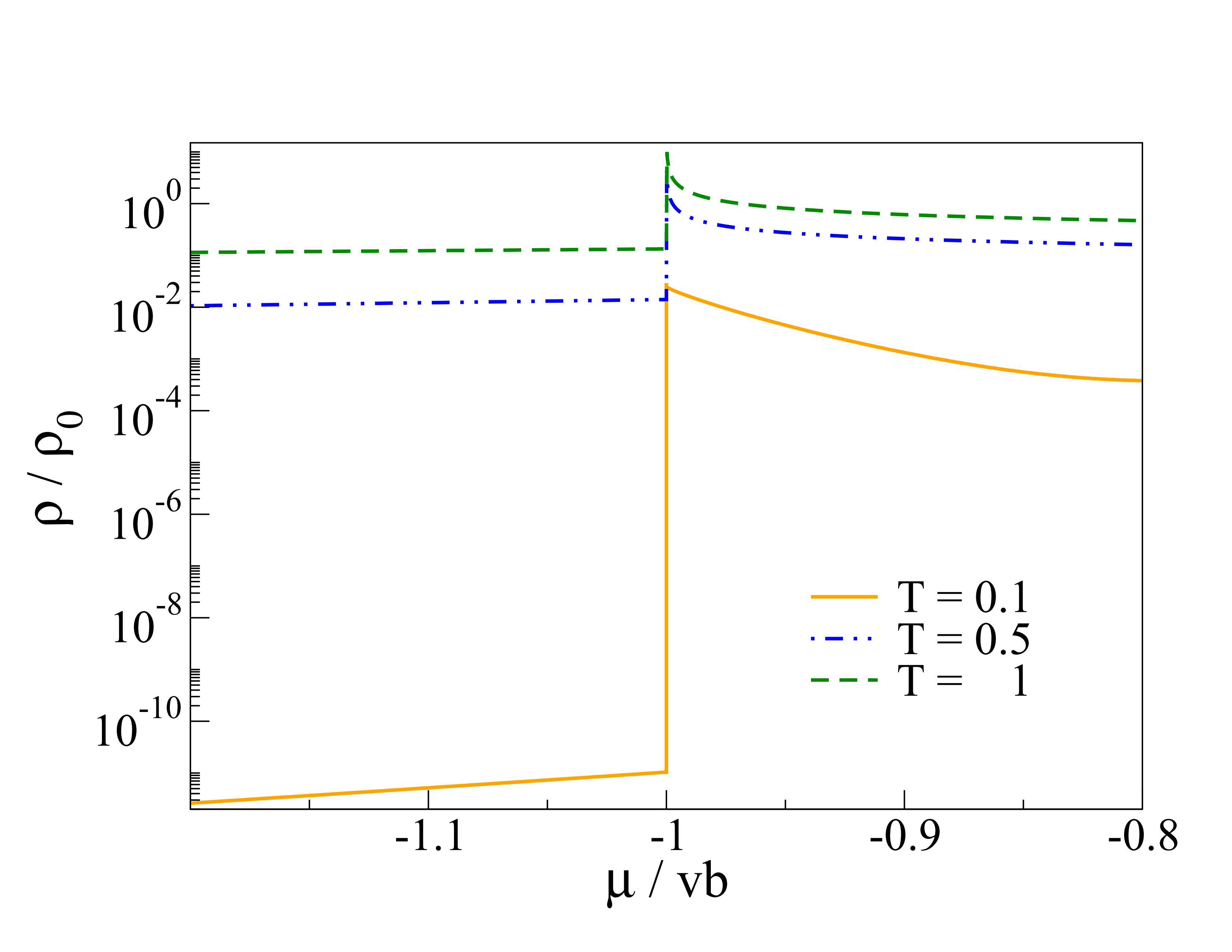}
\caption{Resistivity $\rho$ (in units of $\rho_0$) vs chemical potential $\mu$ (in units of $vb$)
for a two-valley band with dispersion \eqref{toymodel}
across the transition point $\mu=\mu_c$ separating regions with $N=1$ and $N=2$ pairs of Fermi points.
The curves are obtained from Eq.~\eqref{rhocont} with ${\cal I}_{k_+,k_-} = 0.5$.
Results are shown for different temperatures (in units of $T_b=2c_L b$) on a logarithmic scale for $\rho$.
}
\label{fig12}
\end{figure}

To demystify the jump-like behavior of the resistivity reported in Sec.~\ref{sec4b}, 
we consider a toy model for a two-valley subband with the dispersion relation ($v=b=1$)
\begin{equation}\label{toymodel}
    \varepsilon_k=-\left|k^2-1\right|,
\end{equation}
and analyze how the resistivity depends on the chemical potential $\mu<0$. 
For $\mu>\mu_c=-1$, there are $N=2$ pairs of Fermi points, $\pm k_\pm$, 
with $k_\pm=\sqrt{1\pm |\mu|}$ and respective Fermi velocities $v_\pm = 2\sqrt{1\pm |\mu|}$.
On the other hand, for $\mu<\mu_c$, there is only a single pair ($N=1$), $\pm k_F$, with $k_F=k_+$ 
and $v_F=v_+$. 
Therefore, according to Eq.~\eqref{rhocont}, for $\mu>\mu_c$, the dominant resistivity 
contribution comes from intra-bs processes with Bloch-Gr\"uneisen temperature $T_{\rm intra-bs}=c_L(k_+-k_-)$. 
For $\mu<\mu_c$, instead, only inter-bs processes are possible and the relevant 
Bloch-Gr\"uneisen temperature is $T_{\rm inter-bs}=2c_L k_+$. 
The resistivity is thus parametrically larger on the $N=2$ side since intra-bs processes are then possible, 
which are not available on the $N=1$ side. This gives rise to a large jump of the resistivity when 
$\mu$ crosses the critical value $\mu=\mu_c$, as ilustrated in Fig.~\ref{fig12}.

%

We then conclude that the abrupt resistivity changes observed in Sec.~\ref{sec4b} originate 
from transitions between one and two pairs of Fermi points  within a two-valley band.  

\begin{thebibliography}{0}%
\makeatletter
\providecommand \@ifxundefined [1]{%
 \@ifx{#1\undefined}
}%
\providecommand \@ifnum [1]{%
 \ifnum #1\expandafter \@firstoftwo
 \else \expandafter \@secondoftwo
 \fi
}%
\providecommand \@ifx [1]{%
 \ifx #1\expandafter \@firstoftwo
 \else \expandafter \@secondoftwo
 \fi
}%
\providecommand \natexlab [1]{#1}%
\providecommand \enquote  [1]{``#1''}%
\providecommand \bibnamefont  [1]{#1}%
\providecommand \bibfnamefont [1]{#1}%
\providecommand \citenamefont [1]{#1}%
\providecommand \href@noop [0]{\@secondoftwo}%
\providecommand \href [0]{\begingroup \@sanitize@url \@href}%
\providecommand \@href[1]{\@@startlink{#1}\@@href}%
\providecommand \@@href[1]{\endgroup#1\@@endlink}%
\providecommand \@sanitize@url [0]{\catcode `\\12\catcode `\$12\catcode
  `\&12\catcode `\#12\catcode `\^12\catcode `\_12\catcode `\%12\relax}%
\providecommand \@@startlink[1]{}%
\providecommand \@@endlink[0]{}%
\providecommand \url  [0]{\begingroup\@sanitize@url \@url }%
\providecommand \@url [1]{\endgroup\@href {#1}{\urlprefix }}%
\providecommand \urlprefix  [0]{URL }%
\providecommand \Eprint [0]{\href }%
\providecommand \doibase [0]{http://dx.doi.org/}%
\providecommand \selectlanguage [0]{\@gobble}%
\providecommand \bibinfo  [0]{\@secondoftwo}%
\providecommand \bibfield  [0]{\@secondoftwo}%
\providecommand \translation [1]{[#1]}%
\providecommand \BibitemOpen [0]{}%
\providecommand \bibitemStop [0]{}%
\providecommand \bibitemNoStop [0]{.\EOS\space}%
\providecommand \EOS [0]{\spacefactor3000\relax}%
\providecommand \BibitemShut  [1]{\csname bibitem#1\endcsname}%
\let\auto@bib@innerbib\@empty
\end{thebibliography}%


\begin{thebibliography}{99}

\bibitem{Armitage2017} 
N. P. Armitage, E. J. Mele, and A. Vishwanath, 
Rev. Mod. Phys.  {\bf 90}, 015001 (2018).

\bibitem{Hasan2017} 
M. Z. Hasan, S. Y. Xu, I. Belopolski, and S. M. Huang, 
Annu. Rev. Condens. Matt. Phys. {\bf 8}, 289 (2017).

\bibitem{Burkov2018}
A. A. Burkov, 
Annu. Rev. Condens. Matt. Phys. {\bf 9}, 359 (2018).

\bibitem{Lv2021} 
B. Q. Lv, T. Qian, and H. Ding, 
Rev. Mod. Phys. {\bf 93}, 025002 (2021).

\bibitem{Ong2021} 
N. P. Ong and S. Liang, 
Nat. Rev. Phys.  {\bf 3},  394 (2021).

\bibitem{Zhang2021a}
C. Zhang, Y. Zhang, H.-Z. Lu, X. C. Xie, and F. Xiu, 
Nature Rev. Phys. {\bf 3}, 660 (2021).

\bibitem{Spivak2016}
B. Z. Spivak and A. V. Andreev, 
Phys. Rev. B {\bf 93}, 085107 (2016).

\bibitem{Baireuther2016} 
P. Baireuther, J. A. Hutasoit, J. Tworzydlo, and C. W. J. Beenakker, 
New J. Phys. {\bf 18}, 045009 (2016). 

\bibitem{Baireuther2017} 
P. Baireuther, J. Tworzydlo, M. Breitkreiz, I. Adagideli,   and C. W. J. Beenakker, 
New J. Phys. {\bf 19}, 025006 (2017). 

\bibitem{Igarashi2017}
A. Igarashi and M. Koshino, 
Phys. Rev. B {\bf 95}, 195306 (2017).

\bibitem{Erementchouk2018}
M. Erementchouk and P. Mazumder, 
Phys. Rev. B {\bf 97}, 035429 (2018).

\bibitem{Kaladzhyan2019}
V. Kaladzhyan and J. H. Bardarson, 
Phys. Rev. B {\bf 100}, 085424 (2019).

\bibitem{Sukhachov2020}
P. O. Sukhachov, M. V. Rakov, O. M. Teslyk, and E. V. Gorbar, 
Ann. Phys. (Berlin) {\bf 532}, 1900449 (2020).

\bibitem{Gorbar2016}
E. V. Gorbar, V. A. Miransky, I. A. Shovkovy, and P. O. Sukhachov, 
Phys. Rev. B {\bf 93}, 235127 (2016).

\bibitem{Breitkreiz2019} 
M.  Breitkreiz and P. W. Brouwer, 
Phys. Rev. Lett. {\bf 123}, 066804 (2019). 

\bibitem{Nguyen2020}
T. Nguyen, F. Han, N. Andrejevic, R. Pablo-Pedro, A. Apte, Y. Tsurimaki \emph{et al.}, 
Phys. Rev. Lett. {\bf 124}, 236401 (2020).

\bibitem{Zhang2020}
K. Zhang, X. Pang, T. Wang, F. Han, S.-L. Shang, N. T. Hung, A. R. T. Nugraha, Z.-K. Liu, 
M. Li, R. Saito, and S. Huang, 
Phys. Rev. B {\bf 101}, 014308 (2020).

\bibitem{Hein2020}
P. Hein, S. Jauernik, H. Erk, L. Yang, Y. Qi, Y. Sun, C. Felser, and M. Bauer, 
Nat. Commun. {\bf 11}, 2613 (2020).

\bibitem{Osterhoudt2021}
G. B. Osterhoudt, Y. Wang, C. A. C. Garcia, V. M. Plisson, J. Gooth, C. Felser, P. Narang, and K. S. Burch,
Phys. Rev. X {\bf 11}, 011017 (2021).

\bibitem{Rinkel2019}
P. Rinkel, P. L. S. Lopes, and I. Garate, 
Phys. Rev. B {\bf 99}, 144301 (2019).

\bibitem{Resta2018}
G. Resta, S.-T. Pi, X. Wan, and S. Y. Savrasov, 
Phys. Rev. B {\bf 97}, 085142 (2018).

\bibitem{Voit1987}
J. Voit and H. J. Schulz, 
Phys. Rev. B {\bf 34}, R7429 (1986).

\bibitem{Bockelmann1990}
U. Bockelmann and G. Bastard, 
Phys. Rev. B {\bf 42}, 8947 (1990).

\bibitem{Shik1993}
A. Y. Shik and L. J. Challis, 
Phys. Rev. B {\bf 47}, 2082 (1993).

\bibitem{Mickevicius1993}
R. Mickevicius and V. Mitin, 
Phys. Rev. B {\bf 48}, 17194 (1993).

\bibitem{Gurevich1995}
V. L. Gurevich, V. B. Pevzner, and K. Hess, 
Phys. Rev. B {\bf 51}, 5219 (1995).

\bibitem{Gurevich1995b}
V. L. Gurevich, V. B. Pevzner, and E. W. Fenton, 
Phys. Rev. B {\bf 51}, 9465 (1995).

\bibitem{Seelig2005}
G. Seelig, K. A. Matveev, and A. V. Andreev, 
Phys. Rev. Lett. {\bf 94}, 066802 (2005).

\bibitem{Yurkevich2013}
I. V. Yurkevich, A. Galda, O. M. Yevtushenko, and I. V. Lerner, 
Phys. Rev. Lett. {\bf 110}, 136405 (2013).

\bibitem{Vazifeh2013}
M. M. Vazifeh and M. Franz, 
Phys. Rev. Lett. {\bf 111}, 027201 (2013).

\bibitem{Okugawa2014}
R. Okugawa and S. Murakami, 
Phys. Rev. B {\bf 89}, 235315 (2014).

\bibitem{Bovenzi2018}
N. Bovenzi, M. Breitkreiz, T. E. O'Brien, J. Tworzydlo, and C. W. J. Beenakker, 
New J. Phys. {\bf 20}, 023023 (2018).

\bibitem{Burrello2019}
M. Burrello, E. Guadagnini, L. Lepori, and M. Mintchev, 
Phys. Rev. B {\bf 100}, 155131 (2019).

\bibitem{Witten2016}
E. Witten, 
Nuovo Cim. Riv. Ser. {\bf 89}, 313 (2016).

\bibitem{Landau7}
L. D. Landau and E. M. Lifshitz, 
\textit{Theory of Elasticity} (Elsevier, 1986).

\bibitem{Landau10}
L. D. Landau and E. M. Lifshitz, 
\textit{Course of Theoretical Physics Vol.~10: Physical Kinetics} (Butterworth-Heinemann, 1981).

\bibitem{Levchenko2020}
A. Levchenko and J. Schmalian, 
Ann. Phys. {\bf 419}, 168218 (2020).

\bibitem{Dorn2020}
K. Dorn, A. De Martino, and R. Egger, 
Phys. Rev. B {\bf 101}, 045402 (2020).

\bibitem{Bardarson2013}
J. H. Bardarson and J. E. Moore, 
Rep. Prog. Phys. {\bf 76}, 056501 (2013). 

\bibitem{Jauregui2016}
L. A. Jauregui, M. T. Pettes, L. P. Rokhinson, L. Shi, and Y. P. Chen, 
Nature Nanotechnol.  {\bf 11}, 345 (2016).

\bibitem{Wang2016}
L.-X. Wang, C.-Z. Li, D.-P. Yu, and Z.-M. Liao, 
Nature Commun. {\bf 7}, 10769 (2016).

\bibitem{Lin2017}
B.-C. Lin, S. Wang, L.-X. Wang, C.-Z. Li, J.-G. Li, D. Yu, and Z.-M. Liao, 
Phys. Rev. B {\bf 95}, 235436 (2017).

\bibitem{Lin2020}
B.-C. Lin, S. Wang, A.-Q. Wang, Y. Li, R.-R. Li, K. Xia, D. Yu, and Z.-M. Liao, 
Phys. Rev. Lett. {\bf 124}, 116802 (2020).

\bibitem{Bagoyan2020}
J. R. Bayogan, K. Park, Z. B. Siu, S. J. An, C.-C. Tang, X.-X. Zhang, M. S. Song, J. Park, 
M. B. A. Jalil, N. Nagaosa, K. Hirakawa, C. Sch\"onenberger, J. Seo, and M. Jung, 
Nanotechnology {\bf 31}, 205001 (2020).

\bibitem{Li2021}
C.-Z. Li, A.-Q. Wang, C. Li, W.-Z. Zheng, A. Brinkman, D.-P. Yu, and Z.-M. Liao, 
Phys. Rev. Lett.  {\bf 126}, 027001 (2021).

\bibitem{Nair2020}
N. L. Nair, M.-E. Boulanger, F. Lalibert{\'e}, S. Griffin, S. Channa, A. Legros, W. Tabis, C. Proust, J. Neaton, L. Taillefer, and J. G. Analytis,
Phys. Rev. B {\bf 102}, 075402 (2020).

\bibitem{Cohn2020}
I. A. Cohn, S. G. Zybtsev, A. P. Orlov, and S. V. Zaitsev-Zotov, 
JETP Lett. {\bf 112}, 88 (2020).

\bibitem{Ramshaw2018} B. J. Ramshaw, K. A. Modic, A. Shekhter, Y. Zhang, E.-A. Kim, P. J. W. Moll,
M. D. Bachmann, M. K. Chan, J. B. Betts, F. Balakirev, A. Migliori, N. J. Ghimire, E. D. Bauer,
F. Ronning, and  R. D. McDonald, 
Nature Comm. {\bf 9}, 2217 (2018).

\bibitem{NIST}
\emph{NIST Digital Library of Mathematical Functions}, edited by F. W. J. Olver, A. B. Olde Daalhuis, D. W. Lozier, 
B. I. Schneider, R. F. Boisvert, C. W. Clark, B. R. Miller, and B. V. Saunders, 
available at http://dlmf.nist.gov/, Release 1.0.16 of 2017-09-18.

\bibitem{Peng2016} 
B. Peng, H. Zhang, H. Shao, H. Lu, D. W. Zhang, and H. Zhu, 
Nano Energy {\bf 30}, 225 (2016).

\bibitem{shojaei2021}
I. A. Shojaei, S. Pournia, C. Le, B. R. Ortiz, G. Jnawali, F.-C. Zhang, S. D. Wilson, 
H. E. Jackson, and L. M. Smith, 
Sci. Rep. {\bf 11}, 8155 (2021).

\bibitem{Nazarov2009}
Yu. V. Nazarov and Ya. M. Blanter, 
\textit{Quantum Transport} (Cambridge University Press, 2009).

\bibitem{Kedem2020}
Y. Kedem, E. J. Bergholtz, and F. Wilczek, 
Phys. Rev. Res. {\bf 2}, 043285 (2020).

\bibitem{Gomez2021}
A. G{\'o}mez and L. Urrutia, 
preprint arXiv:2106.15062.

\bibitem{Ilan2020}
R. Ilan, A. G. Grushin, and D. I. Pikulin, 
Nat. Rev. Phys. {\bf 2}, 29 (2020).

\end{thebibliography}
\end{document}